\documentclass[%
reprint,
superscriptaddress,
nofootinbib,
amsmath,
amssymb,
aps,
floatfix,
showkeys,
]{revtex4-2}

\usepackage[colorlinks,citecolor=blue,linkcolor=blue,anchorcolor=blue,filecolor=blue,
urlcolor=blue]{hyperref}

\usepackage{graphicx,color,xcolor}
\usepackage{amsfonts,amsmath}
\usepackage{amssymb,ulem}
\usepackage{dcolumn}
\usepackage{bm,lipsum}
\usepackage[utf8]{inputenc}
\usepackage{subeqnarray}
\usepackage{array}
\usepackage{epsfig,hyperref}
\usepackage{morefloats}
\usepackage{relsize}
\usepackage{url}

\usepackage{comment}

\usepackage{orcidlink}

\begin{document}

\preprint{ }

\title{Effects of the Symmetry energy slope on the exotic content of the neutron stars}

\author{Luiz L. Lopes}

\affiliation{ Centro Federal de Educac\~{a}o Tecnol\'{o}gica de Minas Gerais Campus VIII; CEP 37.022-560, Varginha - MG - Brazil\\}
% \mbox{$^2$Depto de F\'{\i}sica - CFM - Universidade Federal de Santa Catarina  Florian\'opolis - SC - CP. 476 - CEP 88.040 - 900 - Brazil }}

\date{\today}

\begin{abstract}
By varying the symmetry energy slope ($L$), I investigate how the exotic content within the interiors of neutron stars changes and how it affects both macroscopic and microscopic quantities. Using two different parametrizations (L3$\omega\rho$ and BigApple), and three different possibilities about the neutron star core (nucleons+hyperons, nucleons+deltas, nucleons+hyperons+deltas), I show that, for the models analyzed in this work, changing the slope barely changes the amount of hyperons, but it can strongly suppress the $\Delta$ resonances for large values of $L$. I also show that, in general, the presence of exotic content will be more evident for lower values of $L$ than for large ones. Differences and similarities between the two parametrizations are also analyzed. 
\end{abstract}

\maketitle

\section{Introduction}

The Pauli exclusion principle states that two or more fermions cannot simultaneously occupy the same quantum state within a system that obeys the laws of quantum mechanics~\cite{Sakuraibook}.
Neutron stars, conversely, are cold objects (T = 0 K is a valid approximation) formed in principle by neutrons, protons, and leptons in a chemical equilibrium configuration.
As the density increases towards the star's core, the energy of the nucleons also increases. At sufficiently high densities, the increase in the nucleon chemical potentials, driven by Pauli blocking, together with the constraints of chemical equilibrium, renders it energetically favorable to replace part of the nucleonic population by more exotic, non-nucleonic baryonic degrees of freedom. The more usual exotic baryons that can be present at the neutron stars' core are the hyperons~\cite{Glenbook}. The study of hyperons in neutron stars is an old subject~\cite{BETHE1974,GLENDENNING1982} but is still an active field nowadays~\cite{Lopes2022ApJ,Sedrakian2023,Choro2024}. Indeed, two QHD-based  papers~\cite{Dapo2010,lopesnpa} suggest that hyperons are inevitable, and changes in parameterizations or even the inclusion of additional repulsive fields (such as the $\phi$ meson) cannot suppress the hyperon threshold.  In the same sense, the inevitability of hyperons is not exclusive of QHD-based models; ab initio approaches also indicate that hyperons are always present, as can be seen in Refs.~\cite{Tong2025APJ,Mohammad2026APJ}.

Beyond hyperons, other exotic non-nucleonic degrees of freedom that can be present at the neutron star core are the $\Delta$ resonances. Although the $\Delta$ particles are more massive than the $\Lambda^0$ hyperon, they present a much more attractive potential depth. Indeed, modern results point out that the $\Delta$ potential depths are stronger than the nucleon itself, $U_{\Delta}\approx$ -90 to -100 MeV~\cite{Bodek2020}. In recent years, several authors have explored the effect of such a high potential depth~\cite{Oliveira2007,KOLO2017,lopesPRD,Kauan2022PRC}, the main conclusion being that not only are the $\Delta$ resonances present, but the $\Delta^{-}$ is the first exotic degree of freedom to appear.

Concerning nuclear properties, the authors of two different review articles (ref.~\cite{Dutra2014,Micaela2017}), were able to constrain five nuclear quantities at saturation density: saturation density itself $(n_0)$, effective nucleon mass $(M^{*}_N/M_N)$ the binding energy per baryon $(B/A)$, (in) compressibility ($K$), and symmetry energy $(S_0)$. A sixth quantity, the symmetry energy slope, or simply the slope ($L$), is still a matter of debate. In the early 2010s, most studies pointed to a relatively low value for $L$. For instance, in refs.~\cite{Paar2014,Steiner2014,Lattimer2013}  upper limits of 54.6, 61.9, and 66 MeV, respectively, were suggested. However, the situation has changed in the last couple of years, and new experiments have pointed to a significantly higher upper limit.
For example, in a study on the spectra of pions in intermediate energy collisions, an upper limit of 117 MeV was obtained~\cite{pions}, while in one of the PREX II analyses~\cite {PREX2} an upper limit of 143 MeV was suggested.
All these conflicting results have been well summarized in a recent paper~\cite{Tagami2022}: the CREX group points to a slope in the range 0 $<~L~<$ 51 MeV, while PREX II results point to 76 MeV $<~L~<$ 165 MeV. The CREX and PREX II results do not overlap. It is a huge problem that must be solved.

In addition to the constraints related to nuclear physics at the saturation density, some constraints come from neutron star observations, which are related to the behaviour of the nuclear matter at densities far above the saturation point. For example, the mass and radius of the PSR J0740+6620 lies at $M_\odot = 2.08 \pm 0.07$ and 11.41 km $<~R~<$ 13.70 km respectively~\cite{Riley2021}. Concerning the canonical 1.4 $M_\odot$ star, two NICER teams have pointed to a limit of $13.85 ~\mathrm{km}$ \cite{Riley:2019yda} and $14.26 ~\mathrm{km}$ \cite{Miller:2019cac}. These results were refined in ref.~\cite{Miller2021}  to 11.80 km $<R_{1.4}< 13.10$ km. A weaker and more conservative constraint coming from state-of-the-art theoretical results at low and high baryon density points to an upper limit of $R_{1.4}$ $<$ 13.6 km~\cite{Annala2018PRL}. Still on the canonical star, another important quantity and constraint is the so-called dimensionless tidal deformability parameter $\Lambda$. The gravitational wave observations by LIGO/VIRGO in the GW170817 event put constraints on the dimensionless tidal parameter of the canonical star $\Lambda_{1.4}<800$~\cite{Abbott2017}. This result was then refined in ref.~\cite{AbbottPRL}, to  $70<\Lambda_{1.4}.<580$. A more conservative constraint was suggested in ref.~\cite{Li2021}, pointing out the range $133<\Lambda_{1.4}<686$ from a Bayesian analysis.

In this work, I study the influence of the symmetry energy slope on the exotic content of neutron stars. To accomplish that task, I fix the five well-known parameters at the saturation density and run over the slope from 44 MeV up to 92 MeV by adding the non-linear $\omega-\rho$ coupling as presented in the IUFSU model~\cite{IUFSU,Rafa2011,dex19jpg}. It is also possible to obtain $L$ above 92 MeV with the help of the scalar-isovector $\delta$ meson~\cite{KUBIS1997,Liu2002,Lopes2014BJP}. However,   three recent papers~\cite{lopescesar,lopes2024PRC,lopes2024PRCb} indicate that higher values of the slope are in disagreement with some constraints coming from neutron stars' observations, once it predicts the radii for the canonical stars outside the limits inferred by the NICER observations~\cite{Riley:2019yda,Miller:2019cac}, produces a hadron-quark phase transition very close or even below the saturation density, and enable direct URCA process to occur for neutron stars with masses below 1.35$M_{\odot}$~\cite{klahn2006}. 

From the microscopic point of view, I study the particle population and the square of the speed of sound $v_s^2$. Such a quantity is well known to be sensitive to the onset of new degrees of freedom and possesses information about its equation of state (EOS). In relation to the macroscopic results, I study the mass-radius relation as well as the dimensionless tidal parameter. Moreover, to study the model-dependency of the results, I use two different parameterizations of the quantum hadrodynamics (QHD)~\cite{Serot_1992,Miyatsu2013}. I select a soft EOS, the L3$\omega\rho$~\cite{Lopes2022CTP}, and a stiff one, the BigApple~\cite{BigApple}. Nevertheless, even the ``soft'' EOS must be sufficiently stiff to support neutron stars with masses $M \gtrsim 2.0\,M_\odot$, even in the presence of hyperons. Therefore, the stiff EOS must be significantly stiffer; otherwise, the predictions would be too similar. 
{Consequently, the BigApple parametrization is in tension with some constraints coming from astronomical observation, especially the dimensionless tidal parameter of the canonical star. It is, nevertheless, used here mainly as an illustrative stiff-limit case to explore model dependence.} Both chosen parameterizations satisfy all five constraints from nuclear matter at saturation density discussed in Refs.~\cite{Dutra2014,Micaela2017}, while exhibiting distinct behavior at high densities and, consequently, different neutron-star properties.

\section{Formalism and parametrizations}

The extended version of the QHD~\cite{Serot_1992}, which includes the $\omega\rho$ non-linear coupling~\cite{IUFSU,Rafa2011,dex19jpg}  has the following Lagrangian density in natural units:
\begin{eqnarray}
\mathcal{L}_{QHD} =  \sum_B \bar{\psi}_B[\gamma^\mu(\mbox{i}\partial_\mu  - g_{BB\omega}\omega_\mu   - g_{BB\rho} \frac{1}{2}\vec{\tau} \cdot \vec{\rho}_\mu) \nonumber \\
- (M_N - g_{BB\sigma}\sigma )]\psi_B  -U(\sigma) + \frac{1}{2} m_v^2 \omega_\mu \omega^\mu    \nonumber   \\
  + \frac{1}{2}(\partial_\mu \sigma \partial^\mu \sigma - m_s^2\sigma^2) + \frac{\xi g_\omega^4}{4}(\omega_\mu\omega^\mu)^2   \nonumber \\ - \frac{1}{4}\Omega^{\mu \nu}\Omega_{\mu \nu} + \Lambda_{\omega\rho}(g_{\rho}^2 \vec{\rho^\mu} \cdot \vec{\rho_\mu}) (g_{\omega}^2 \omega^\mu \omega_\mu)  \nonumber \\
 + \frac{1}{2} m_\rho^2 \vec{\rho}_\mu \cdot \vec{\rho}^{ \; \mu} - \frac{1}{4}\bf{P}^{\mu \nu} \cdot \bf{P}_{\mu \nu} + \mathcal{L}_\phi,  \nonumber \\   \label{s1} 
\end{eqnarray}
where the $\psi_B$  represent both the baryonic  Dirac field of the baryon octet and of the decuplet, with mass $M_B$. From a rigorous point of view, the members of the baryon decuplet should be described by the Rarita-Schwinger Lagrangian density. Nonetheless,  the resulting equation of motion can be written compactly as a Dirac equation with the same energy eigenvalue as done in ref.~\cite{lopesPRD}.  The $\sigma$, $\omega_\mu$,  and $\vec{\rho}_\mu$ are the mesonic fields.  The $g's$ are the Yukawa coupling constants that simulate the strong interaction,   $m_s$, $m_v$, and $m_\rho$ are
 the masses of the $\sigma$, $\omega$,  and $\rho$ mesons respectively. The $\xi$ is related to the self-interaction of the $\omega$ meson, while the $\Lambda_{\omega\rho}$ is a non-linear coupling between the $\omega$-$\rho$ mesons and controls the symmetry energy and its slope~\cite{dex19jpg}. 
The $U(\sigma)$ is the self-interaction term introduced in ref.~\cite{Boguta} to fix the incompressibility:

\begin{equation}
U(\sigma) =  \frac{\kappa M_N(g_{\sigma} \sigma)^3}{3} + \frac{\lambda(g_{\sigma}\sigma)^4}{4} \label{sbog} ,
\end{equation} 
and $\mathcal{L}_\phi$ is related  the strangeness hidden $\phi$ vector
meson, which couples only with the hyperons ($Y$), not affecting the
properties of symmetric  nuclear matter:

\begin{equation}
\mathcal{L}_\phi = g_{YY \phi}\bar{\psi}_Y(\gamma^\mu\phi_\mu)\psi_Y + \frac{1}{2}m_\phi^2\phi_\mu\phi^\mu - \frac{1}{4}\Phi^{\mu\nu}\Phi_{\mu\nu} , \label{EL3} 
\end{equation}.

Furthermore, leptons are added as free fermions to account for the chemical stability
and electrical charge neutrality. Therefore, we have the following.

\begin{equation}
 \mu_B =  \mu_n - Q_B\mu_e, \quad \mu_e = \mu_\mu,\quad  \sum_f Q_f n_f = 0 \label{ce}
\end{equation}
where $\mu_B$ and $Q_B$ are the chemical potential and the electric charge of the baryon $B$, 
$\mu_n$, $\mu_e$, and $\mu_\mu$ are the chemical potentials of the neutron, electron, and muon, respectively. The sum in $f$ implies over all the fermions, and $n$ is the number density.

The equation of state (EoS) is then obtained in mean-field approximation (MFA) by calculating the components of the energy-momentum tensor. The detailed calculation of the EOS in the mean field approximation can be found in ~\cite{Serot_1992,IUFSU,Lopes2014BJP,Lopes2022ApJ,Miyatsu2013,Glenbook,lopes2023ptep,LopesUNIVERSE2025} and the references therein.

For each parametrization, except for the slope, all the other five quantities ($n_0$, $M^{*}_N/M_N$, $B/A$, $K$, $S_0$)  are fixed. This implies that, except for $g_\rho$ and $\Lambda_{\omega\rho}$, all other parameters of the model are also fixed. In Tab.~\ref{TL1}, I present the fixed parameters of the L3$\omega\rho$ and BigApple models, the prediction of the five fixed quantities, and the constraints coming from ref.~\cite{Dutra2014,Micaela2017}. In Tab.~\ref{T2}, I present the values of $g_\rho$ and $\Lambda_{\omega\rho}$ for each model and for four different slope values.

 As can be seen from Tab.~\ref{T2}, for $L = 92$ MeV within the L$3\omega\rho$ parametrization, the non-linear coupling $\Lambda_{\omega\rho}$ is zero. Therefore, the results will reflect not only the slope, but also the effects of a null $\Lambda_{\omega\rho}$, as this term at high densities limits the growth of the $\rho$ field.

\begin{center}
\begin{table}[h]
\begin{center}
\begin{tabular}{c|cc|c}
\hline
    & L3$\omega\rho$~\cite{Lopes2022CTP}   &BigApple~\cite{BigApple} & Constraints~\cite{Dutra2014,Micaela2017} \\
\toprule
 $n_0$  (fm$^{-3}$) & 0.156  & 0.155 &  0.148 - 0.170  \\
 $K$ (MeV) &256 &229  & 220 - 260  \\
$M^{*}_N/M_N$  & 0.69 &  0.61   & 0.6 - 0.8    \\ 
$B/A$ (MeV) & 16.2 &  16.3 & 15.8 - 16.5   \\
$S_0$ (Mev)  &31.7 & 31.3   & 28.6 - 34.4     \\
 \hline
 $M_N$ (MeV) &938.93 &  939.0  & -   \\
 $m_\sigma$ (MeV) &512 &   492.7& - \\
 $m_\omega$  (MeV)&783 &  782.5  & -       \\
 $m_\rho$ (MeV)&770 &  763  &  - \\
 $m_\phi$ (MeV)& 1020 & 1020 & - \\
 $g_\sigma$  &9.029 &  9.670  & -    \\
$g_\omega$  &10.597 &   12.316  & -  \\
$\kappa$ &0.00414 &   0.00277 & -   \\
$\lambda$ &-0.00390 &-0.00362 & - \\
$\xi$    & -  & 0.00012 & -   \\
\hline 
\end{tabular}
\caption{Parameter sets used in this work and corresponding saturation properties and constraints. } \label{TL1}
\end{center}
\end{table}
\end{center}

%------------------------------------------------------
\begin{center}
\begin{table}[h]
\begin{center}
\begin{tabular}{ccccc}
\toprule
Model &  $L$ (MeV) & $g_\rho$ & $\Lambda_{\omega\rho}$  \\
\toprule
L3$\omega\rho$ & 44 & 11.310 &  0.0515 \\
 %\hline
L3$\omega\rho$ & 60 & 9.685  &  0.0344 \\
%  \hline
L3$\omega\rho$ & 76& 8.638 &  0.0171        \\
% \hline
L3$\omega\rho$ &92 &  7.863  &  0  \\
\hline
BigApple & 44 & 12.846 &  0.0440 \\
 %\hline
BigApple & 60 & 10.000  &  0.0310 \\
%  \hline
BigApple & 76& 8.509 &  0.0181        \\
% \hline
BigApple &92 &  7.567  &  0.0057  \\
\toprule
\end{tabular}
\caption{Parameters for four fixed values of the symmetry energy slope within six different parametrizations.} 
\label{T2}
\end{center}
\end{table}
\end{center}

Finally, in the presence of hyperons and $\Delta$ resonances, it is also necessary to fix the coupling constants for these particles. There is very little experimental data related to the strength of hyperons and the $\Delta$' interaction with nuclear matter. Some of them are the potential depths. The hyperon potential depths are known with a satisfactory degree of accuracy. Here I use $U_{\Lambda}$ = -28 MeV, $U_{\Sigma}$ = +30 MeV, and $U_{\Xi}$ = -4 MeV~\cite{Potentials2000,LQCD}. Nevertheless, recent studies point out that the potential of the $\Sigma$'s are less repulsive~\cite{HaidenbauerEPJA2023}, while the potential of the $\Xi$'s are more attractive~\cite{FriedmanEPJWC2022,FRIEDMAN2025PLB}. However, those papers also point out that these large values can be reduced substantially by the repulsive three-body force contributions.

The $\Delta$  potential depth, on the other hand, does not share the precision of the hyperons; however, it is well-accepted that it is more attractive than the nucleon. Here, I use $U_{\Delta}$ = -90 MeV, which is in agreement with the recent analysis~\cite{Bodek2020,KOLO2017}.

Since the potential depth of a baryon $B$ is given by:

\begin{equation}
U_B = g_{BB\omega}\omega_0 - g_{BB\sigma}\sigma_0,     \label{depth}
\end{equation}
 there are several sets of coupling constants that produce the same potential. One way to reduce the number of free parameters is by using symmetry group arguments. Using the SU(3) group, the coupling constants of the vector mesons with all the baryons can be fixed in terms of just one free parameter, $\alpha_V$ (see refs.~\cite{lopesPRD,lopes2023ptep} and the references therein for a detailed discussion). To produce massive stars even with the appearance of exotic particles, here I use $\alpha_V$ = 0.5, which gives us:

 \begin{eqnarray}
  \frac{g_{\Delta\Delta\omega}}{g_{NN\omega}} &=&   1.285, \quad   \frac{g_{\Delta\Delta\rho}}{g_{NN\rho}}  = 1.000, \quad  \frac{g_{\Delta*\Delta*\rho}}{g_{NN\rho}} =3.0, \nonumber \\ 
  \frac{g_{\Lambda\Lambda\omega}}{g_{NN\omega}} &=&   0.714, \quad   \frac{g_{\Lambda\Lambda\phi}}{g_{NN\omega}}  = -0.808, \quad  \frac{g_{\Lambda\Lambda\rho}}{g_{NN\rho}} =0.0, \nonumber \\
   \frac{g_{\Sigma\Sigma\omega}}{g_{NN\omega}} &=& 1.000, \quad  \frac{g_{\Sigma\Sigma\phi}}{g_{NN\omega}} = -0.404, \quad \frac{g_{\Sigma\Sigma\rho}}{g_{NN\rho}}  = 1.0, \nonumber \\
   \frac{g_{\Xi\Xi\omega}}{g_{NN\omega}} &=& 0.571, \quad  \frac{g_{\Xi\Xi\phi}}{g_{NN\omega}} = -1.010, \quad \frac{g_{\Xi\Xi\rho}}{g_{NN\rho}}  = 0.0, \nonumber \\
   \frac{g_{\Sigma\Lambda\rho}}{g_{NN\rho}} &=& 0.577, \label{couplings}
 \end{eqnarray}
where the$\Delta$ and $\Delta^{*}$ are related to the isospin projection ($T_3$) equals to $\pm 1/2$ and $ \pm 3/2$ respectively, as discussed in ref.~\cite{lopesPRD} Furthermore, the $\Delta's$ resonances does not couple to the  strangeness hidden $\phi$ meson. Finally, $g_{BB\sigma}$ is model-dependent and can be obtained from Eq.~\ref{depth} for the desired potential.

It is worth emphasizing that the results are inherently model dependent. Varying $\alpha_V$ modifies the strength of the vector couplings and, as shown in Ref.~\cite{lopesPRD}, for a given baryon $B$, larger values of $g_{BB\omega}$ lead to a reduced population of that species. Similarly, the results depend on the uncertain value of the $\Delta$ potential depth. As demonstrated in Ref.~\cite{Kauan2022PRC}, deeper potentials favor the appearance of $\Delta$ baryons and increase their population. Nevertheless, in the present study, no qualitative changes are expected since variations in $L$ only affect the $\rho$ meson couplings, as well as the $\Lambda_{\omega\rho}$ parameter.

\section{Microscopic Results - Particle population}

I now study how different values of the slope affect the population of exotic particles on neutron stars. I used three different possibilities for the neutron star core: hyperonic matter, consisting of hyperons (Y) and nucleons (N); $\Delta$ matter, with nucleons and $\Delta$ resonances (D); and $\Delta$ admixed hyperonic matter, consisting of NYD matter. 
Also, instead of plotting the total population of particles for fixed values of $L$, here I show the amount of the desired particle for different values of $L$. This allows us to see the differences in one particle population even when the influence of the slope is small.

Furthermore, the slope not only affects the population of exotic particles. The proton fraction is also strongly dependent on the slope at low and high densities. The proton fraction for six different parameterizations with different values of $L$ can be seen in Fig. 1 of ref.~\cite{lopes2024PRCb} and will remain unchanged until the onset of the first exotic particle.

%%%%%%%%%%%%%%%%%
\begin{figure*}[ht]
\begin{tabular}{ccc}
\centering % \begin{center}/\end{center} takes some additional vertical space
\includegraphics[scale=.58, angle=270]{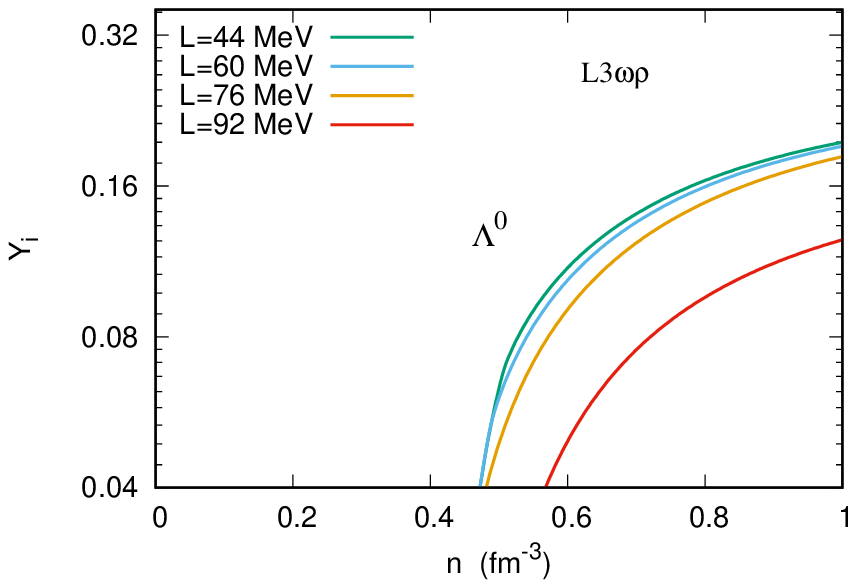} &
\includegraphics[scale=.58, angle=270]{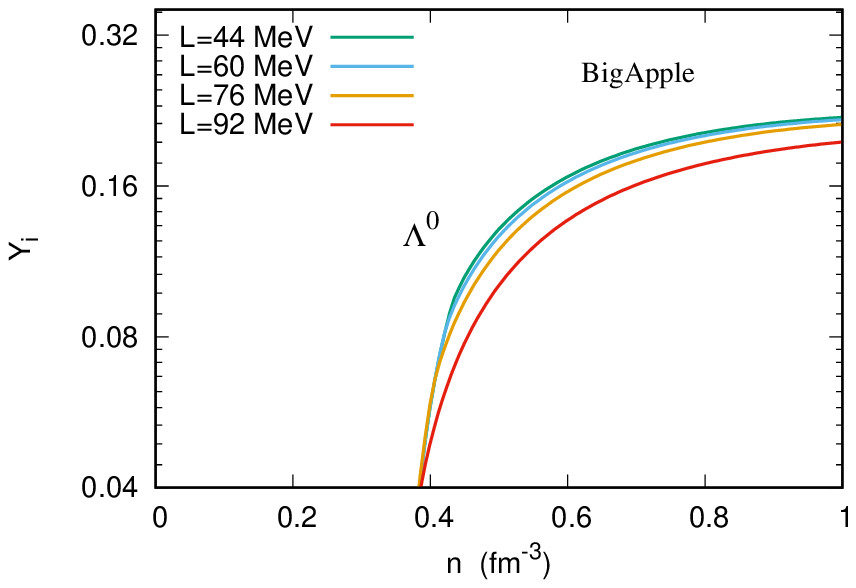} \\
\includegraphics[scale=.58, angle=270]{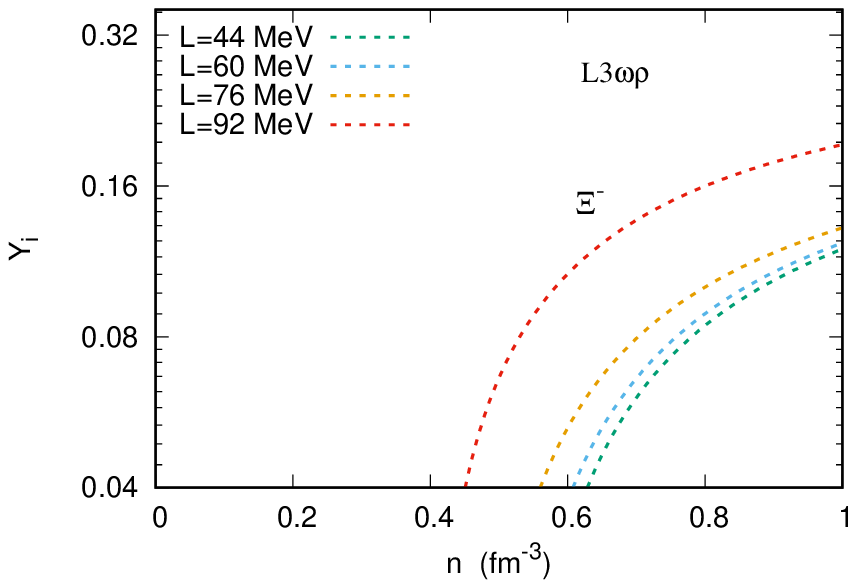} &
\includegraphics[scale=.58, angle=270]{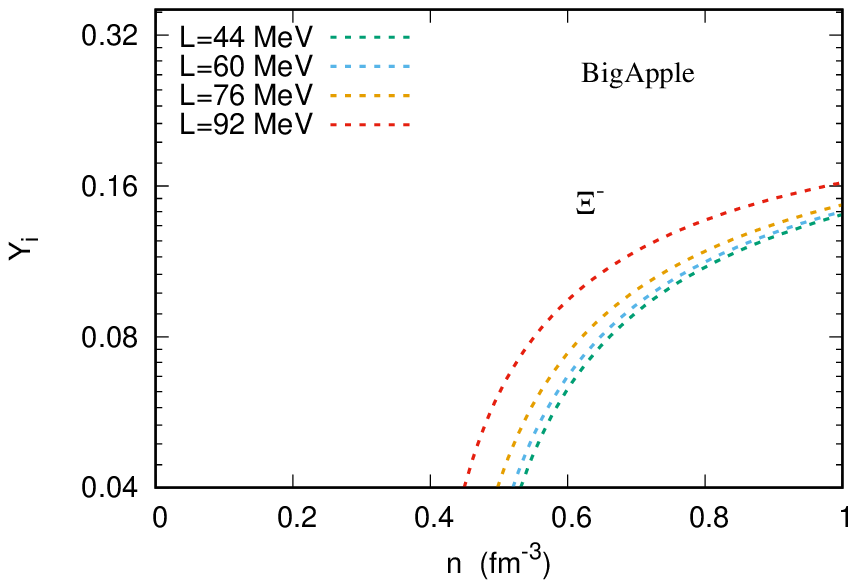}\\
\end{tabular}
\caption{$\Lambda^0$  (above) and $\Xi^{-}$ (below) populations with different values of $L$ with the L3$\omega\rho$ (left) and BigApple (right) parametrizations for NY matter.} \label{YA}
\end{figure*}

\subsection{Hyperonic matter}

I start with the NY matter. Due to the high repulsive potential, combined with a strong coupling to the $\omega$ meson, the $\Sigma$ triplet is not present, independently of the parametrization as well as the slope. The hyperons present are the $\Lambda^0$ and the $\Xi^-$, and the results are shown in Fig.~\ref{YA}. From a qualitative point of view, increasing the slope does not significantly change the onset of $\Lambda^0$, but it does suppress this particle at higher densities. This result is exacerbated for the $L = 92$ MeV in the L3$\omega\rho$, whose onset is pushed to higher densities. The threshold of the $\Xi^-$, on the other hand, is pulled to lower densities as we increase the slope.
From a quantitative point of view, it can be seen that there is little difference between $L$ = 44 and $L$ = 60 MeV. The differences are relevant for $L = 92$ MeV, especially for L3$\omega\rho$, because this parameterization has $\Lambda_{\omega\rho}$ = 0. As discussed in refs.~\cite{lopescesar,lopes2024PRCb}, the $\Lambda_{\omega\rho}$ reduces the $\rho$ field at high densities. Furthermore, for $L = 92$ MeV, within L3$\omega\rho$, the $\Lambda^0$ particle is suppressed at low densities and the onset of the $\Xi^-$ becomes the first exotic particle to appear. This was first noted in ref.~\cite{lopes2023ptep} and it is due to both, $\Lambda_{\omega\rho} = 0$, as well the presence of the mixed term $g_{\Sigma\Lambda\rho}$.

Finally, it can also be pointed out that in the BigApple parameterization, the appearance of exotic particles occurs at lower densities and their population reaches higher values, compared to L3$\omega\rho$. For example, in the BigAppale parameterization, for $L$ = 44 MeV, the threshold $\Lambda^0$ occurs around 0.36 fm$^{-3}$ and the population of $\Lambda^0$ reaches a fraction $Y_\Lambda$ = 0.22 at 1.0 fm$^{-3}$. On the other hand, L3$\omega\rho$ parametrization with the same slope predicts the onset of $\Lambda^0$ around 0.43 $fm^{-3}$, reaching $Y_\Lambda$ = 0.19 at 1.0 fm$^{-3}$.   The only exception is for $\Xi^-$ with $L = 92$ MeV. It appears earlier and reaches high values in L3$\omega\rho$ due to $\Lambda_{\omega\rho} = 0$, and the presence of the mixed term $g_{\Sigma\Lambda\rho}$. In Tab.~\ref{T3}, I show the density where each hyperon reaches the density of 0.01 fm$^{-3}$.

\subsection{$\Delta$ matter}

\begin{figure}[ht]
  \begin{centering}
\begin{tabular}{c}
\includegraphics[width=0.333\textwidth,angle=270]{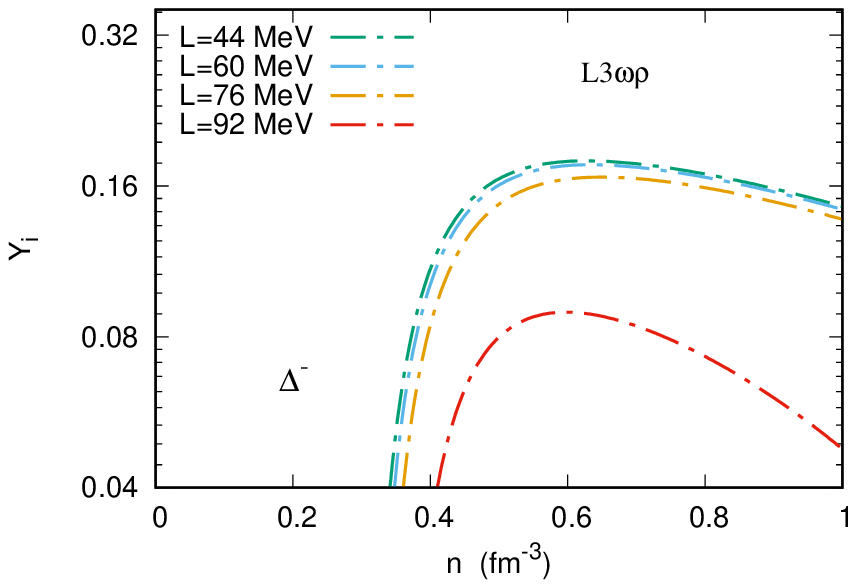} \\
\includegraphics[width=0.333\textwidth,angle=270]{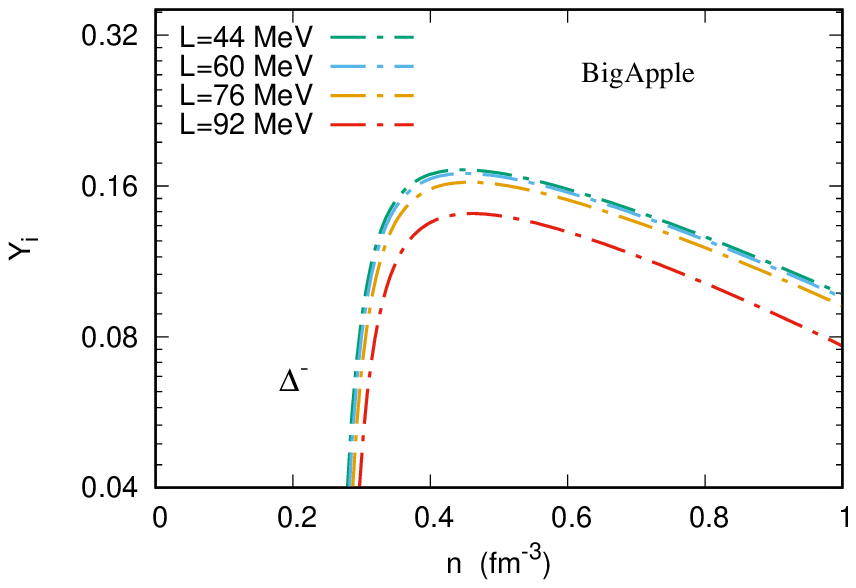} \\
\end{tabular}
\caption{$\Delta^-$ particle for the L3$\omega\rho$ (above) and the BigApple parametrization with different values of $L$ for ND matter.} \label{DA}
\end{centering}
\end{figure}

A second possibility is the onset of $\Delta$ resonances, without the appearance of the hyperons, which I call here ND matter. In ref.~\cite{lopesPRD}, the authors study the onset of $\Delta$ resonances in dense matter for a fixed value of the slope, but changing the value of $\alpha_V$.
They show that usually, $\Delta^-$ and $\Delta^0$ are present. The only exception is for $\alpha_V$ = 0.50, which is the value used in the present work. Therefore, this is an opportunity to investigate whether, by changing the slope or changing the parameterization, $\Delta^0$ can also appear within $\alpha_V$ = 0.50. The results show that $\Delta^0$'s are never present for $\alpha_V$ = 0.50, making $\Delta^-$ the sole exotic particle for ND matter. The $\Delta^-$ population is shown in Fig.~\ref{DA}.

We can see that, analogously to the $\Lambda^0$ particle, increasing the slope pushes the onset of the $\Delta^-$ to higher densities. However, $\Delta^-$ has a different behavior. While $\Lambda^0$ shows a fraction that increases monotonically with density, $\Delta^-$ shows a peak and starts to decrease at higher densities. In the case of ND matter, the presence of the peak happens because $g_{\Delta\Delta\omega}~>g_{NN\omega}$. 
There is an exception again for $L = 92$ MeV, where the $\Delta^-$ fraction is strongly unfavored in the L3$\omega\rho$ due to $\Lambda_{\omega\rho}$ = 0.

From a quantitative point of view, it can be seen that the onset of $\Delta^-$ occurs at even lower densities than $\Lambda^0$. For example, with $L = 44$ MeV, the $\Delta^-$ reach a fraction of 0.01 at 0.27 fm$^{-3}$ for BigApple, and 0.32 fm$^{-3}$ for L3$\omega\rho$. As will be shown in the next section, the onset of exotic particles at such low densities indicates that even the canonical 1.4$M_\odot$ stars present non-nucleonic degrees of freedom at their cores. 

\subsection{Hyperonic $\Delta$ matter.}

%%%%%%%%%%%%%%%%%
\begin{figure*}[ht]
\begin{tabular}{ccc}
\centering % \begin{center}/\end{center} takes some additional vertical space
\includegraphics[scale=.58, angle=270]{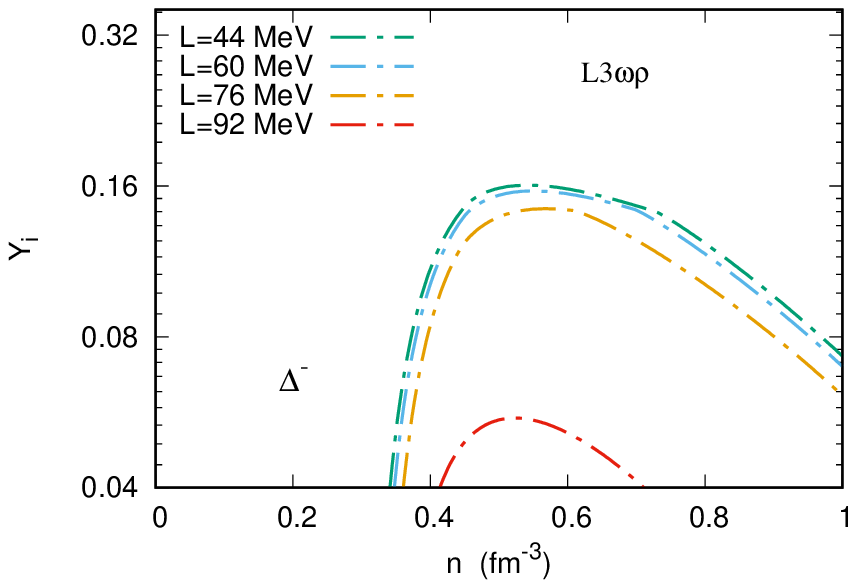} &
\includegraphics[scale=.58, angle=270]{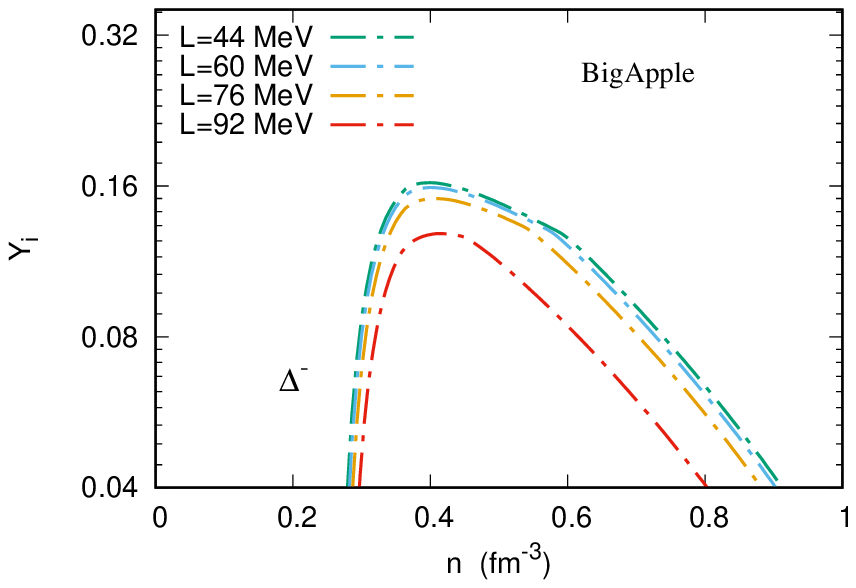} \\
\includegraphics[scale=.58, angle=270]{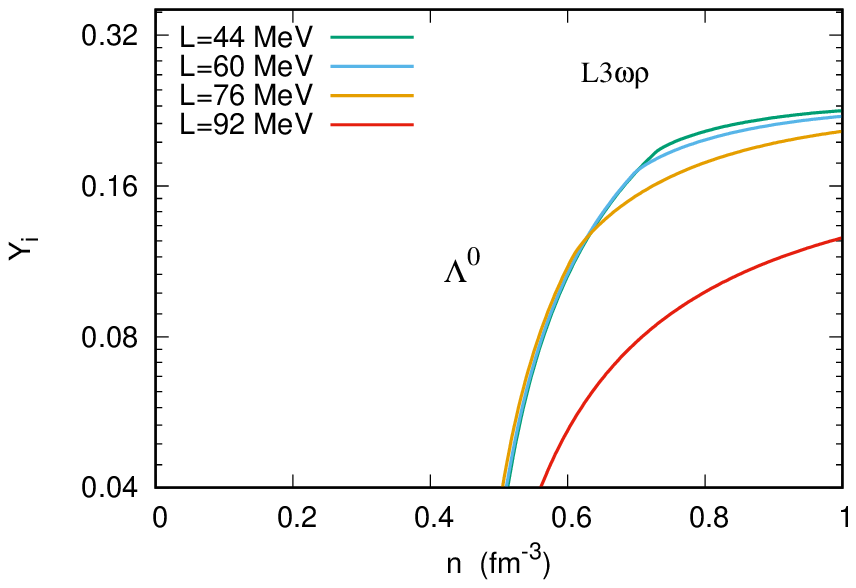} &
\includegraphics[scale=.58, angle=270]{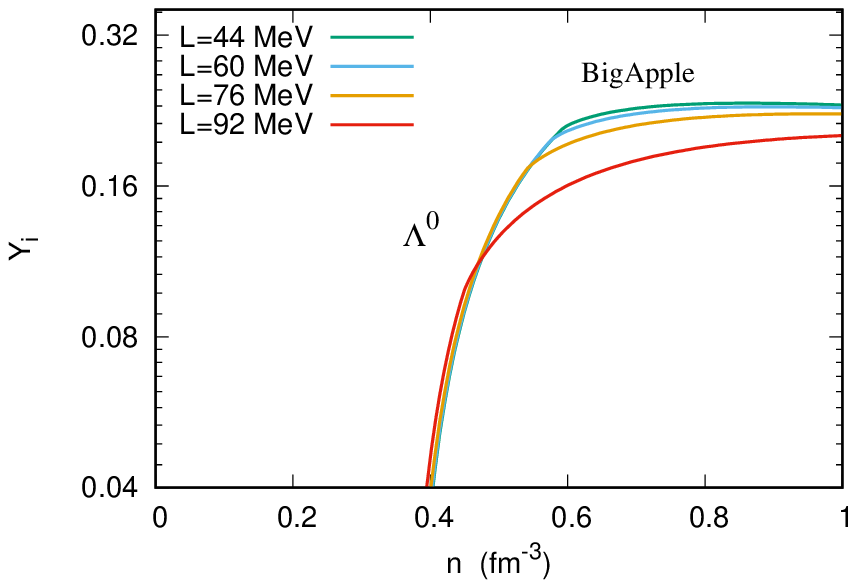}\\
\includegraphics[scale=.58, angle=270]{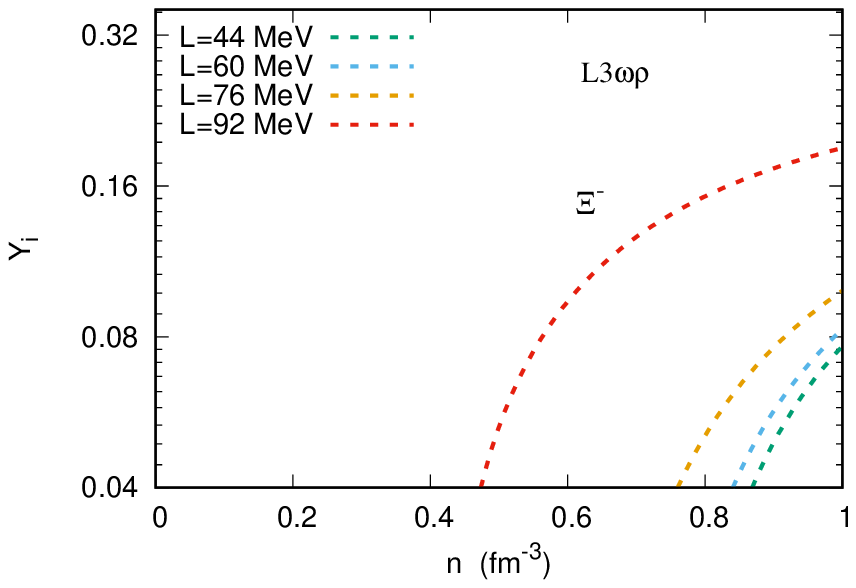} &
\includegraphics[scale=.58, angle=270]{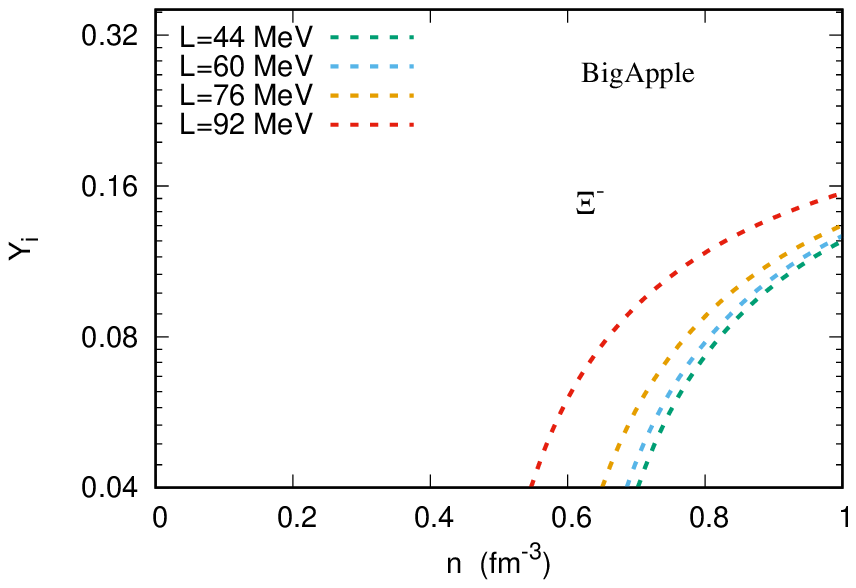}\\
\end{tabular}
\caption{$\Delta^-$ (above), $\Lambda^0$  (midle) and $\Xi^{-}$ (below) populations with different values of $L$ with the L3$\omega\rho$ (left) and BigApple (right) parametrizations for the NYD matter.} \label{YD}
\end{figure*}

The last case I studied is the possibility of both hyperons and $\Delta$ resonances being present at massive neutron star cores, the NYD matter. In this case, three exotic particles are present: $\Delta^-$, $\Lambda^0$, and the $\Xi^-$. The $\Delta^0$ is again absent for all values of $L$, as well as the $\Sigma$ triplet. The results are displayed in Fig.~\ref{YD}.

We notice that, in general, the first exotic particle to appear is the $\Delta^-$, followed by the $\Lambda^0$ and finally the $\Xi^-$. However, the $\Lambda^0$ and the $\Xi^-$ onset are delayed to higher densities. On the other hand, the peak and the consequent decrease with the density presented in the $\Delta^-$ fraction are much more evident in the presence of the hyperons once besides the effect related to $g_{\Delta\Delta\omega}~>g_{NN\omega}$, now the decline of $\Delta^-$ particles is enhanced due to the competition faced with the much less repulsive $\Xi^-$. In the case of the BigApple parametrization, the $\Delta^-$ fraction goes to zero before 1 fm$^{-3}$.
For the $\Lambda^0$'s, we can see that the density of their onset occurs almost independently of the value of $L$. At high density, lower values of $L$ present higher fractions of $\Lambda^0$.

Here again, $L =$ 92 MeV for L3$\omega\rho$ presents an exacerbated behavior {\bf due to $\Lambda_{\omega\rho}$ = 0.} The $\Delta^-$ is strongly suppressed, and its fraction goes to zero for densities above 0.7 fm$^{-3}$. Besides $\Lambda_{\omega\rho}$ = 0, the other factor that makes the quick decrease of the $\Delta^-$ is the competition due to the onset of the $\Xi^-$ at densities only slightly superior to that related to the $\Delta^-$. The very small number of $\Delta$ resonances indicates that the NY and NYD stars will be very similar. All relevant properties are present in Tab.~\ref{T3}.

\section{Microscopic Results - The square of the speed of sound}

I now discuss the quantities related to the equation of state (EOS). Graphically, changing the slope has little effect on the  EOSs. This is the reason that some studies about the symmetry energy do not even present the EOS (as for instance ref.~\cite{Rafa2011}). Others, on the other hand, choose to display the EOS in log scale~\cite{lopescesar}, and even then, the differences are small.

Furthermore, the effects of hyperons and $\Delta$ resonances are also well known in the literature, at least for fixed values of $S_0$ and $L$~\cite{lopesnpa,Bodek2020,Oliveira2007,KOLO2017,lopesPRD,Kauan2022PRC}. The presence of hyperons softens the EOSs, while the presence of $\Delta$'s can soften the EOS at moderate density, but if the coupling of the $\Delta$ resonances with the $\omega$ meson is strong enough, the EOS at high density is as stiff as the pure nucleonic one. Since the EOS has low sensitivity to changes in the slope, I use here the square of the speed of sound:

\begin{equation}
  v_s^2 = \frac{\partial p}{\partial \epsilon}. \label{index}  
\end{equation}

\begin{figure*}[h!]
\begin{tabular}{ccc}
\centering % \begin{center}/\end{center} takes some additional vertical space
\includegraphics[scale=.52, angle=270]{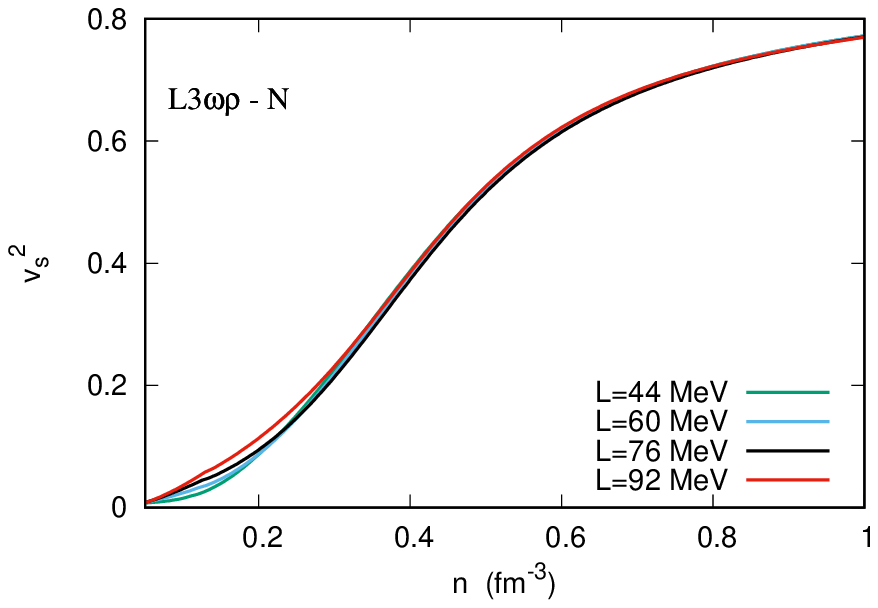} &
\includegraphics[scale=.52, angle=270]{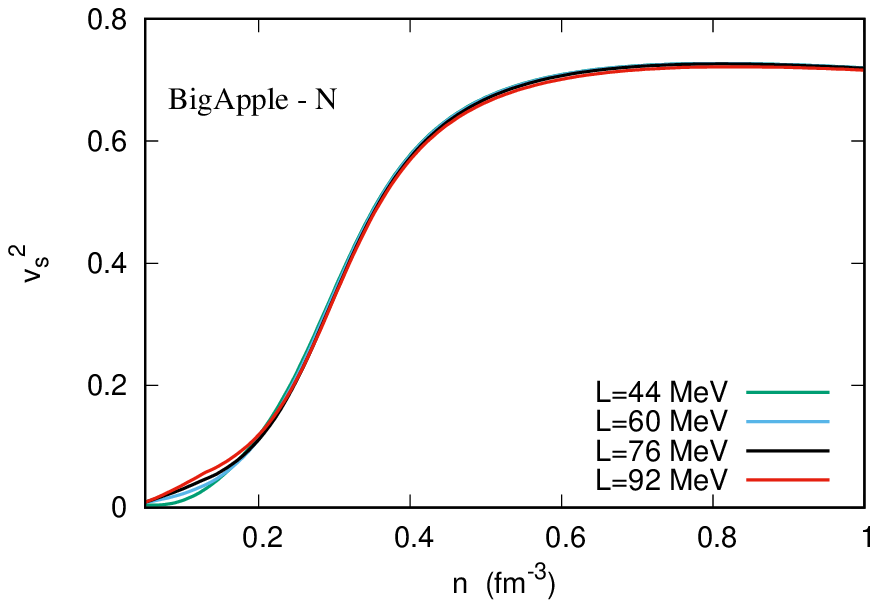} \\
\includegraphics[scale=.52, angle=270]{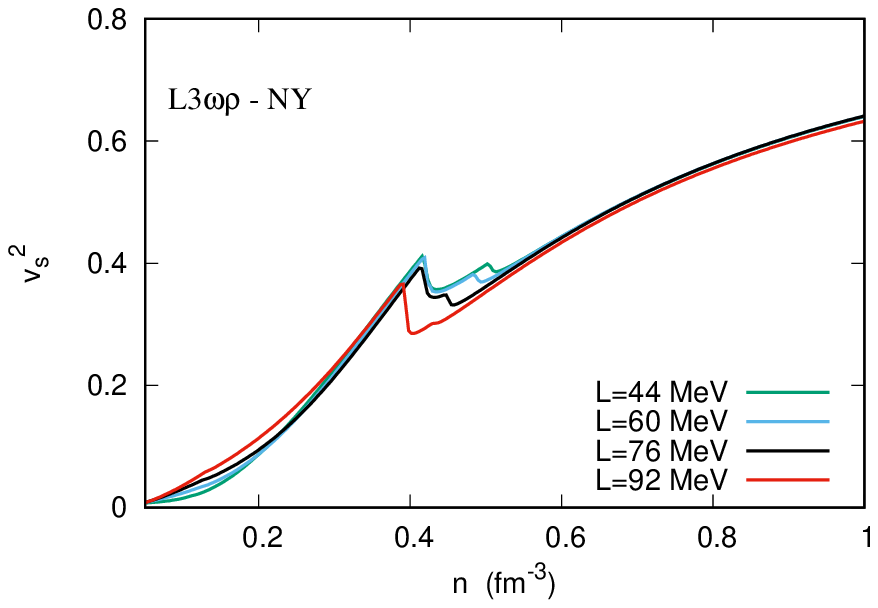} &
\includegraphics[scale=.52, angle=270]{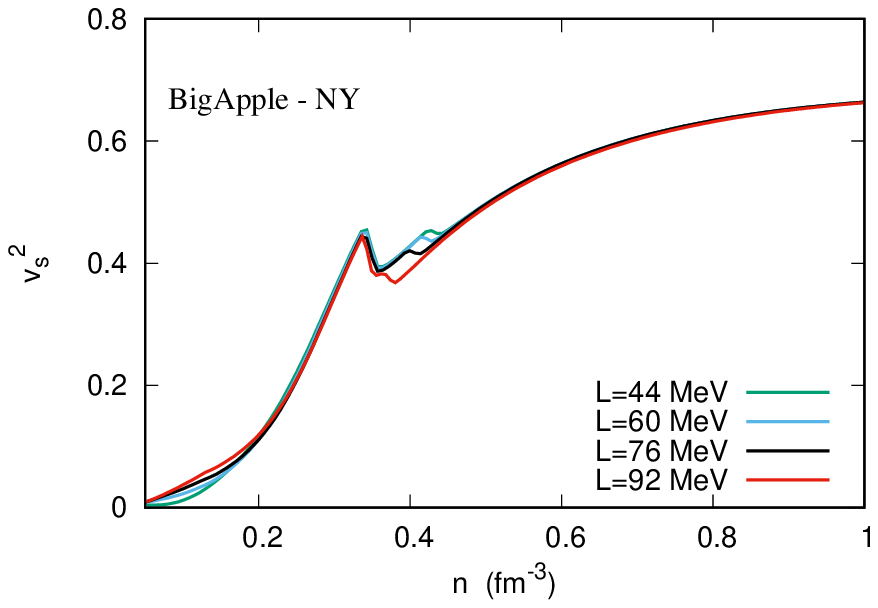}\\
\includegraphics[scale=.52, angle=270]{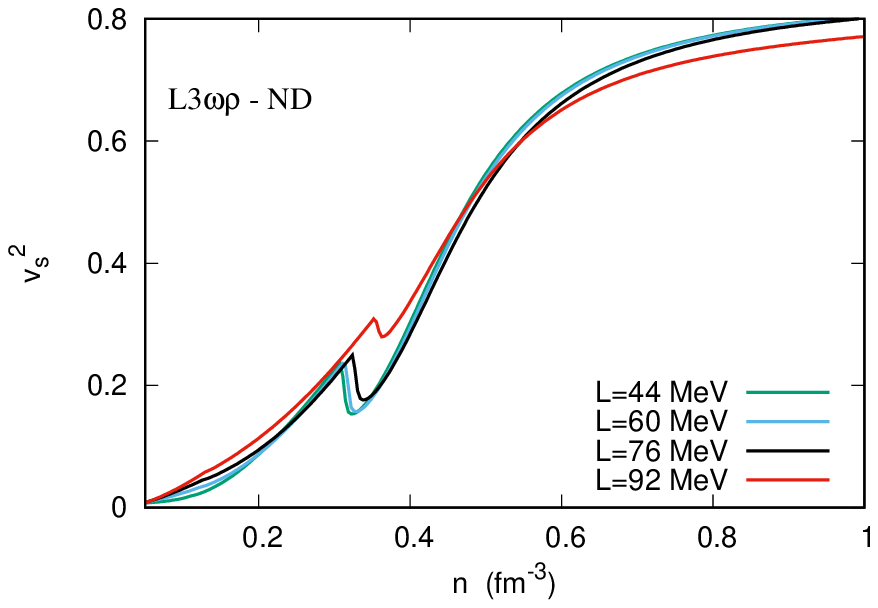} &
\includegraphics[scale=.52, angle=270]{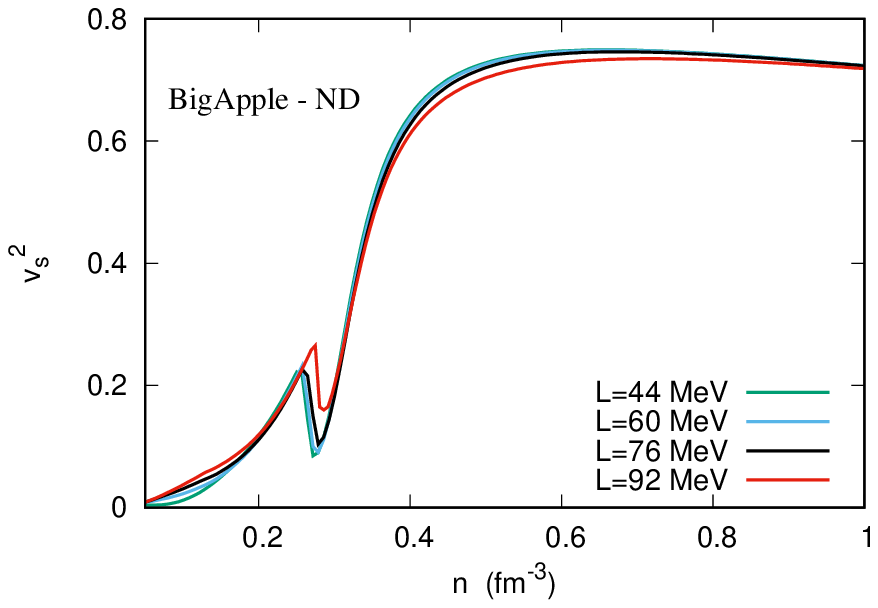}\\
\includegraphics[scale=.52, angle=270]{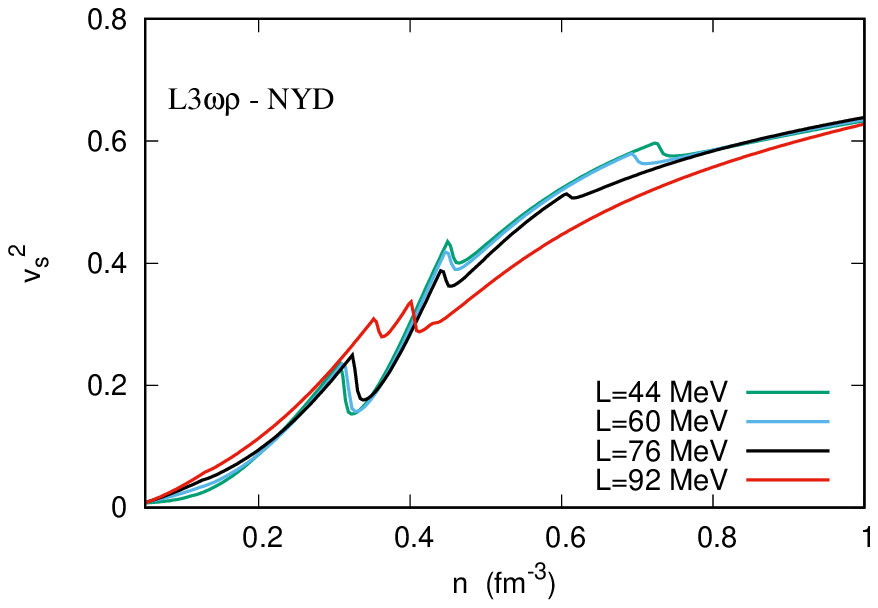} &
\includegraphics[scale=.52, angle=270]{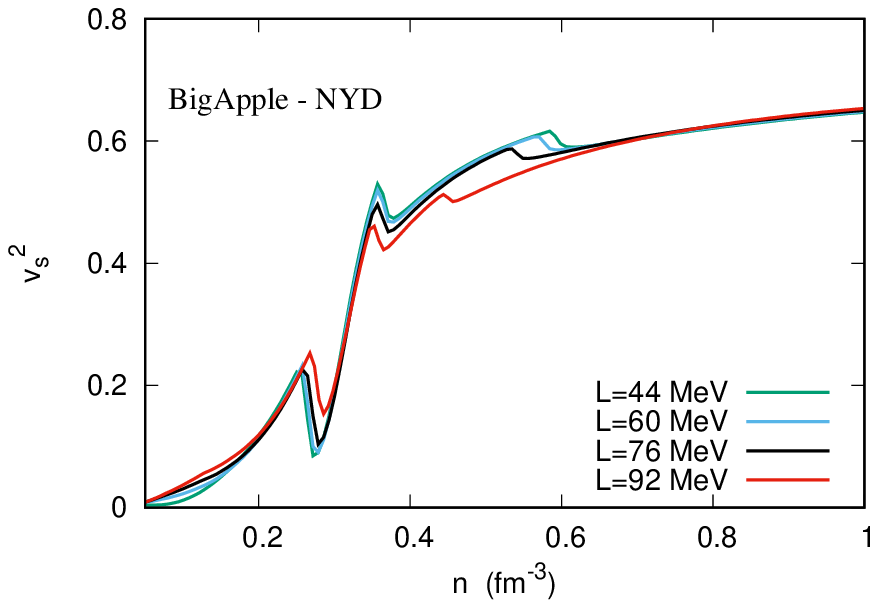}\\
\end{tabular}
\caption{Square of the speed of sound for different stars' configuration for the L3$\omega\rho$ (left) and the BigApple (right) parametrizations.} \label{adi}
\end{figure*}

The square of the speed of sound still presents information about the underlying EOS, as $p = \int v^2_s d\epsilon$. For multicomponent matter, it exhibits drops and kinks at densities coincident with density thresholds of individual components, signaling changes in the matter constitution. The results are presented in Fig.~\ref{adi}.

%%%%%%%%%%%%%%%%%

We can see that the qualitative results, for pure nucleonic matter, the L3$\omega\rho$ present a monotonic growth, while the BigApple has a maximum around 0.7 fm$^{-3}$. 
This is caused by the non-null value of $\xi$ in the Lagrangian for the BigApple (see Tab~\ref{TL1}), which limits the increase of the $\omega$ field. Moreover, the BiggApple presents a steeper curve at moderate densities, indicating a stiffer EOS.

Increasing the slope causes an increasing of the speed of sound at low densities, indicating a stiffer EOS, and consequently a large radii for neutron stars with low masses (see a more complete discussion in Ref.~\cite{LopesUNIVERSE2025}). At high densities, the speed of sound becomes degenerate, indicating similar maximum masses.

For NY matter, the hyperon threshold is clearly identified by a reduction in the speed of sound. In the L3$\omega\rho$ model, a systematic—albeit modest—correlation is observed between the hyperon onset and the slope parameter: larger slopes shift the appearance of the first hyperon to lower densities. A similar trend holds for the second hyperon. Moreover, the density interval between the onset of the first and second hyperons decreases as the slope increases. For the particular case $L = 92$ MeV, the density interval between the onset of the first and second hyperons is so small that the second drop in the speed of sound becomes nearly indistinguishable. This behavior is not solely a consequence of the large slope parameter, but also reflects the fact that this parametrization adopts $\Lambda_{\omega\rho} = 0$. For the BigApple, the first drop to the onset of the $\Lambda^0$ is essentially the same across all values of $L$, but the density interval for the second drop is also reduced as the slope increases. A feature common to both parametrizations is that, at high densities, the speed of sound is reduced relative to pure nucleonic matter, reflecting the well-known softening of the EOS.

For ND matter, a pronounced drop in the speed of sound is observed at the onset of the $\Delta^-$ resonance, followed by a rapid and steep increase, especially for the BigApple parametrization. The exception occurs for $L = 92$ MeV within the L$3\omega\rho$ parametrization, which is associated with $\Lambda_{\omega\rho} = 0$ in this case.
The high-density behaviour shows that the speed of sound is not strongly reduced due to the onset of $\Delta^-$ as in the case of hyperons. Indeed, it is even larger for the L3$\omega\rho$. This happens because $g_{\Delta\Delta\omega}~>g_{NN\omega}$, while such an effect is not present in the BigApple due to the presence of a quartic term in the $\omega$ field, $\xi~\neq 0$. Eventually, it is clear that $L = 92$ MeV produces lower values of $v_s^2$ at high densities for both parametrizations.

Finally, for NYD matter, the behavior reflects a superposition of the features discussed above. The onset of the $\Delta^-$ resonance induces a drop in $v_s^2$, followed by a steep increase. At higher densities, however, the appearance of hyperons softens the EOS, leading to a reduction in the speed of sound. Once again, the case $L = 92$ MeV within the L$3\omega\rho$ parametrization exhibits a markedly different behavior, associated with $\Lambda_{\omega\rho} = 0$.

\section{Macroscopic results}

\subsection{Mass-radius relation}

In this section, I discuss the macroscopic properties of neutron stars (the mass-radius relation and the dimensionless tidal parameter) within different models and different particle populations. The equations of state (EOS) are used as input to obtain both quantities. Therefore, I do not present them here, as they can be indirectly seen in the neutron stars' properties. The mass-radius relation is obtained by solving the TOV equations~\cite{TOV}:

\begin{eqnarray}
 \frac{dp}{dr} = \frac{-GM(r)\epsilon (r)}{r^{2}} \bigg [ 1 + \frac{p(r}{\epsilon(r)} \bigg ]   \bigg  [ 1 + \frac{4\pi p(r)r^3}{M(r)} \bigg ] \nonumber \\ \times \bigg [ 1 - \frac{2GM(r)}{r} \bigg ]^{-1} , \nonumber \\
 \frac{dM}{dr} =  4\pi r^2 \epsilon(r).  \nonumber \\  \label{EL11}
\end{eqnarray}

%%%%%%%%%%%%%%%%%
\begin{figure*}[ht!]
\begin{tabular}{ccc}
\centering % \begin{center}/\end{center} takes some additional vertical space
\includegraphics[scale=.52, angle=270]{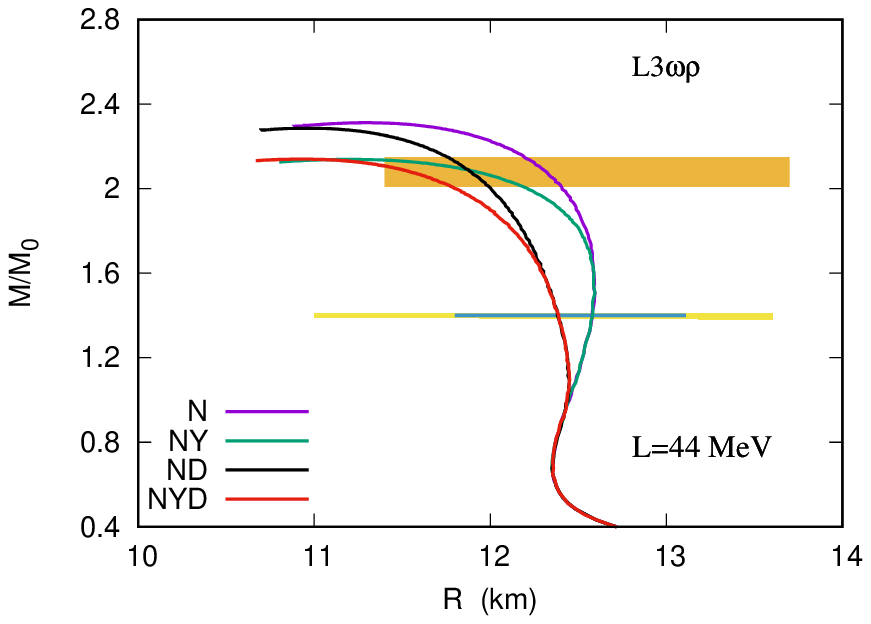} &
\includegraphics[scale=.52, angle=270]{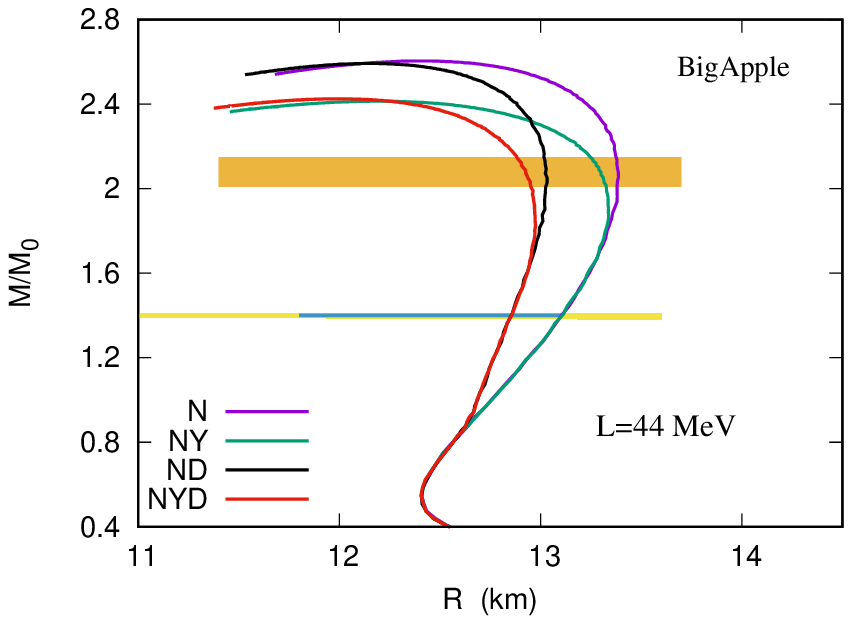} \\
\includegraphics[scale=.52, angle=270]{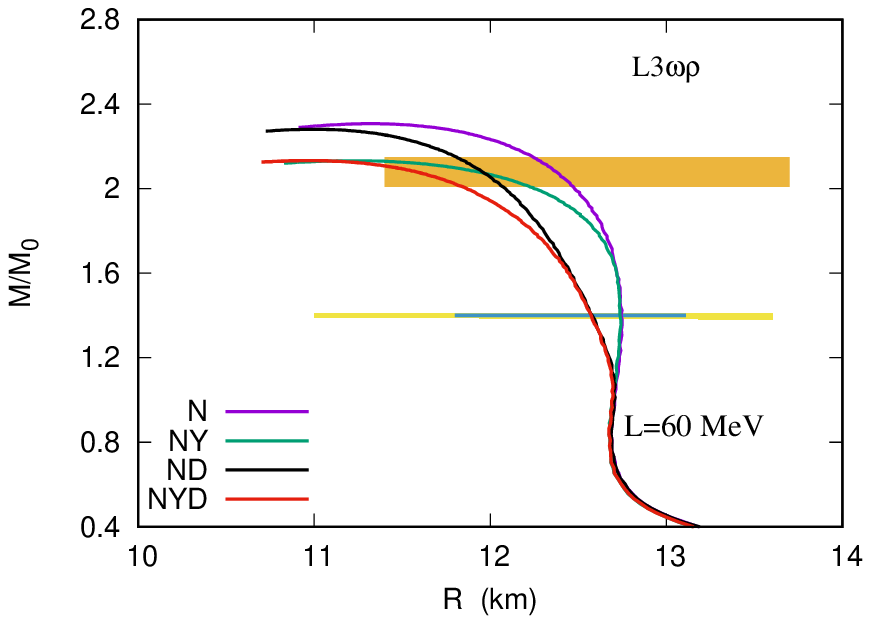} &
\includegraphics[scale=.52, angle=270]{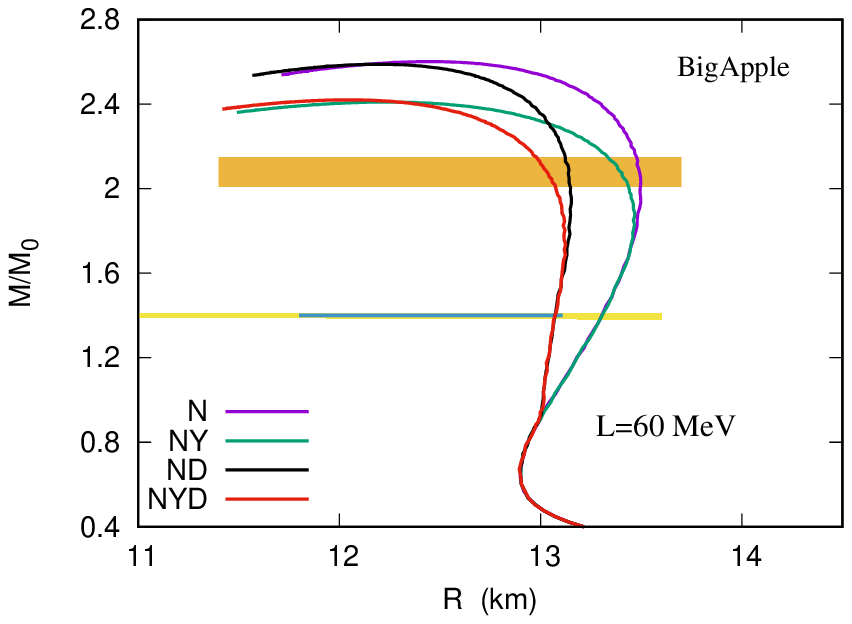}\\
\includegraphics[scale=.52, angle=270]{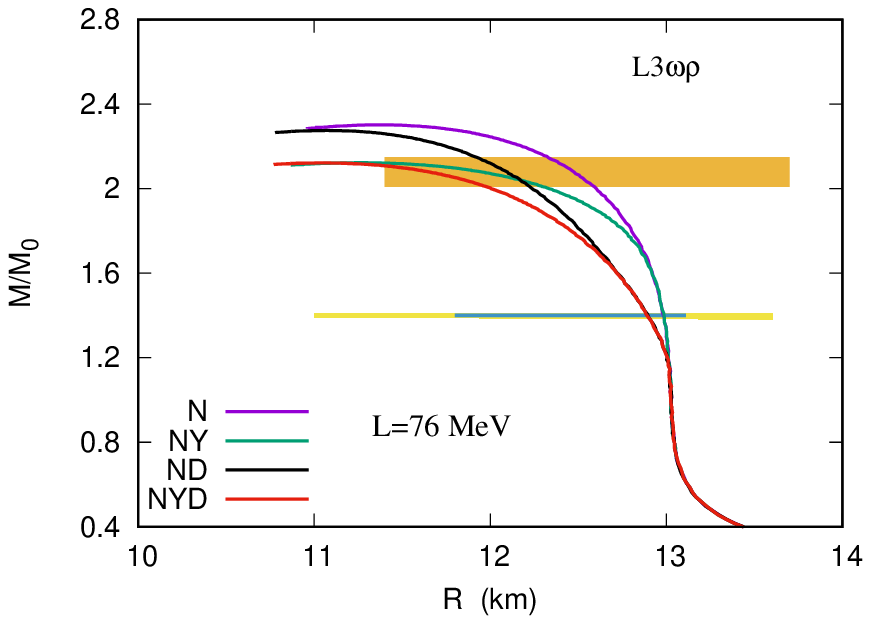} &
\includegraphics[scale=.52, angle=270]{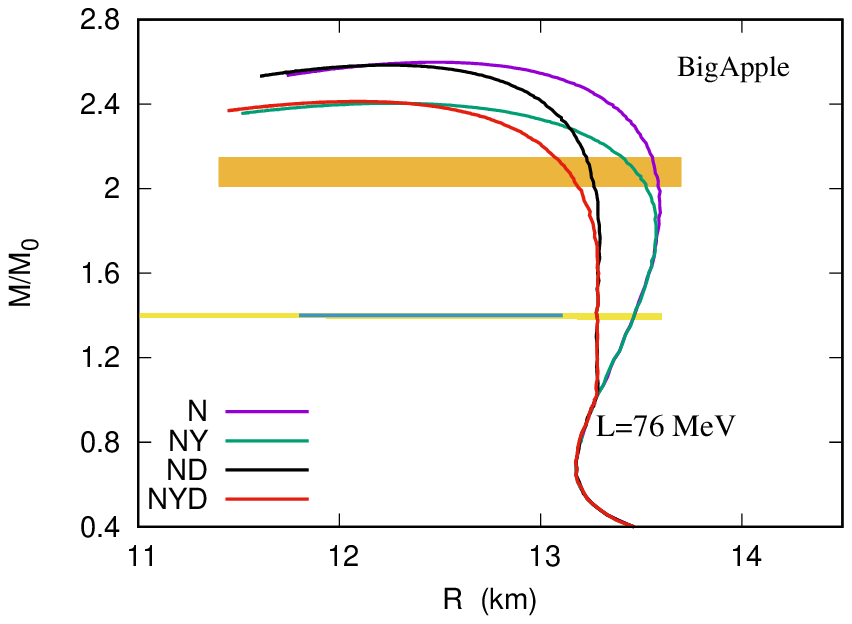}\\
\includegraphics[scale=.52, angle=270]{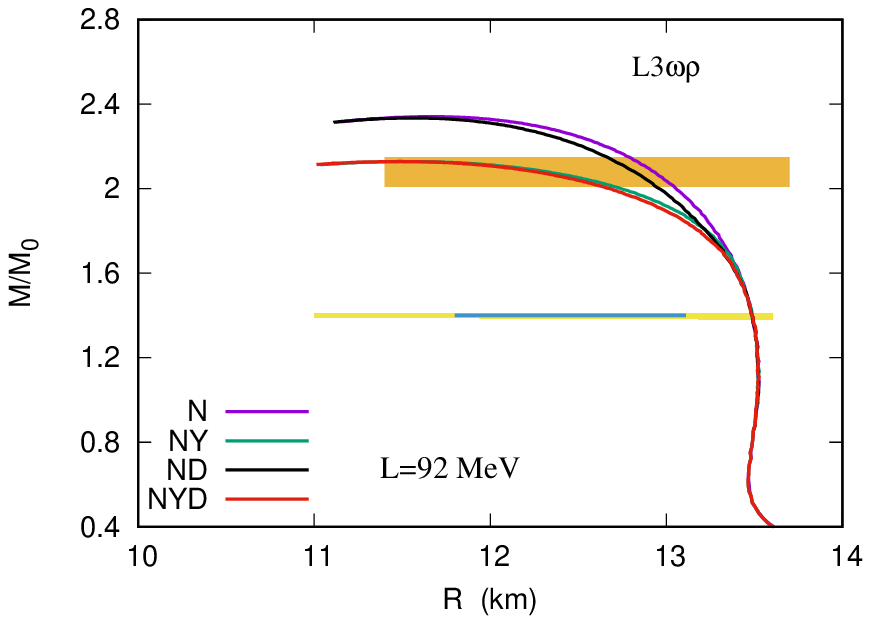} &
\includegraphics[scale=.52, angle=270]{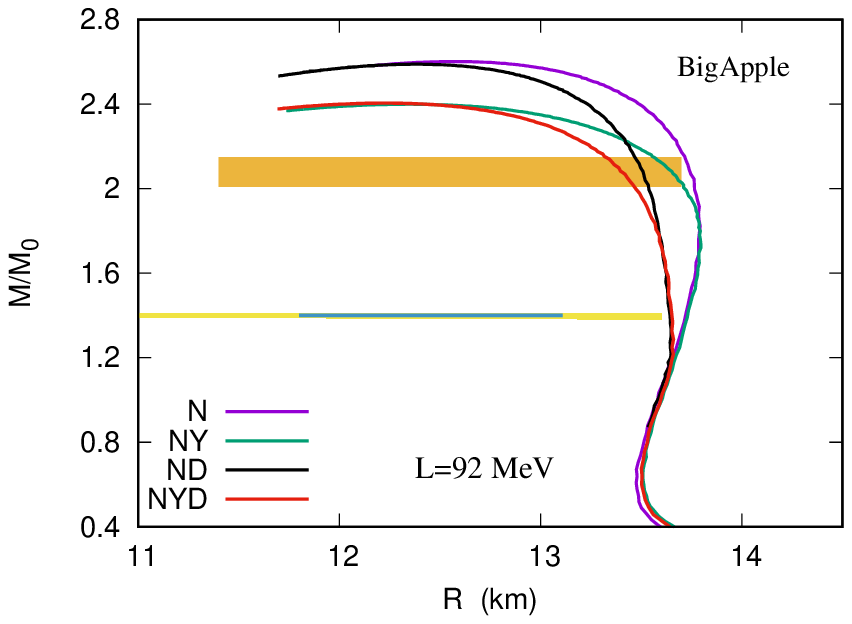}\\
\end{tabular}
\caption{Mass-radius relations for different values of $L$ and different neutron stars' interiors. The hatched areas are constraints as discussed in the text.} \label{tov}
\end{figure*}

To describe the outer and inner crusts of the neutron stars, we utilized the Baym-Pethick-Sutherland (BPS) EOS \cite{BPS} and the Baym-Bethe-Pethick (BBP) EOS \cite{BBP}, respectively.   I use the BPS+BBP EoS up to the density of 0.0089 fm$^{-3}$ for all values of $L$, and from this point on, I use the QHD EoS, as suggested in ref.~\cite{Glenbook}. This procedure is the same as the one done in ref.~\cite{lopescesar,lopes2024PRC,Lopes2024ApJ}.

The effects of the slope of the symmetry energy on the macroscopic properties of standard neutron stars have already been studied in the literature~\cite{Rafa2011,lopes2024PRCb,lopescesar,Lopes2024ApJ}, and it is well known to affect the mass-radius relationship. Therefore, comparing stars with different values of $L$ makes little sense. Instead, what I do in this section is to a given fixed value of $L$, how much the exotic phase influences the macroscopic properties of neutron stars.

Altogether with the mass-radius relations, I also display some observational constraints in Fig.~\ref{tov}. Maybe the more important constraint nowadays is the PSR J0740+6620, with an inferred mass lying at $M = 2.08\pm 0.07~M_\odot$~\ with a radius between 11.41 km $<~R~<$ 13.70 km as suggested in ref.~\cite{Riley2021}. Concerning the canonical 1.4 $M_\odot$ star, two NICER teams have pointed to a limit of $13.85 ~\mathrm{km}$ \cite{Riley:2019yda} and $14.26 ~\mathrm{km}$ \cite{Miller:2019cac}. These results were refined in ref.~\cite{Miller2021}  to 11.80 km $<R_{1.4}< 13.10$ km, which implies a strong bound with an uncertainty of only 5$\%$. A more conservative constraint coming from state-of-the-art theoretical results at low and high baryon density points to an upper limit of $R_{1.4}$ $<$ 13.6 km~\cite{Annala2018PRL}. The range related to the PSR J07040+6620 is presented as an orange-hatched area. The 11.80 km $<R_{1.4}< 13.10$ km range pertaining to the NICER observation is presented as a blue line, and the conservative constraint $R_{1.4}$ $<$ 13.6 km is presented as a yellow line.

It is worth pointing out that the use of a strong and a weak constraint for the canonical star allows me to study the slope $L$ in a large region, $44~$ MeV~$<~L~<92$ MeV. There are, nevertheless, stronger and weaker constraints coming from the NICER X-ray telescope. For instance, Ref~\cite{Mauviard2025APJ} reported an upper limit of only 11.30 km for the canonical pulsar PSR J0614-3329. All of the results presented here are in disagreement with such an extreme constraint, which is so extreme that it is also in disagreement even with the strong constraint of 11.80 km $<R_{1.4}< 13.10$ km presented in Ref.~\cite{Miller2021} and used this work. A not-so-extreme constraint is related to the PSR J0437-4715. Ref.~\cite{Choudhury2024APJL} points to an upper limit $R_{1.4}~<$ 12.30 km. On the other hand, a much weaker constraint was presented in Ref.~\cite{Riley:2019yda}, with a mass lying $M = 1.34^{+0.15}_{-0.16} M_\odot$, this study reports an upper limit of 13.85 km for the  PSR J0030+0451 pulsar.

Now, analyzing Fig.~\ref{tov}, we can first notice that the constraint related to the PSR J0740+6620 pulsar is always satisfied, with the exception of a pure nucleonic star within the BigApple model and with $L = 92$ MeV. In relation to the canonical star, we see that the strong constraint presented by the NICER teams,  11.80 km $<R_{1.4}< 13.10$ km, is satisfied for the $L3\omega\rho$ with the exception of $L = 92$ MeV. But for the BigApple model, it is satisfied only for $L =44$ MeV and $L = 60$ MeV, and only with the condition that $\Delta$'s resonance is present. About the weak constraint,  $R_{1.4}$ $<$ 13.6 km, it is satisfied for all models with the exception of the BigApple within $L = 92$ MeV, which cannot be satisfied even when $\Delta$ resonances are present.

Now I study how different values of $L$ affect the macroscopic properties of exotic stars compared to those of pure nucleonic content.
As in ref.~\cite{lopes2024PRCb,Lopes2024ApJ}, I focus the analysis on the features related to the canonical 1.4$M_\odot$ and the 2.01 $M_\odot$ star.

Let us begin with the canonical 1.4$M_\odot$ stars. 
For such a mass, hyperons are never present, but the $\Delta^-$ particles are always present due to the strong attractive potential $U_\Delta^- = -90$ MeV. In this case, N and NY present the same properties as well as ND and NYD. For the L3$\omega\rho$ model, standard canonical stars (without exotic content) have radii of 12.58 km, 12.74 km, 12.99 km, and 13.48 km for symmetry energy slopes of 44, 60, 76, and 92 MeV, respectively.
The reductions in the radii, $\Delta R$ are, respectively, 0.20 km (1.59$\%$), 0.17 km (1.33$\%$), 0.10 km (0.77$\%$), and 0.00 km, where $\Delta R$ is the difference between the radius of a neutron star with and without exotic content. As can be seen, the reduction in the radii is small. Nevertheless, it is important to point out that there is a correlation between $\Delta R$ and $L$. As $L$ increases, the stars become more and more degenerate and $\Delta R$ decreases. For $L$ = 92 MeV, the $\Delta^-$ fraction is small enough to make any change in the radius of the canonical star. This follows from the fact that this parametrization has $\Lambda_{\omega\rho}=0$.

In the case of the canonical star for the BigApple model, hyperons are again not present; therefore, I only analyze the effect of the $\Delta^-$ particle. The standard stars have radii of 13.03, 13.31, 13.48, and 13.77 km for symmetry energy slopes of 44, 60, 76, and 92 MeV, respectively. The reductions in the radii $\Delta R$ are, respectively: 0.26 km (1.98$\%$), 0.22 km (1.65$\%$), 0.20 km (1.48$\%$), and 0.12 km (0.87$\%$). We can see that this reduction is larger for the BigApple compared with the L3$\omega\rho$. The correlation between $\Delta R$ and $L$ is kept, but $\Delta R$ does not go to zero at $L = 92$ MeV, due to the fact that $\Lambda_{\omega\rho}$ is not zero for BigApple.

Now I analyze 2.01 $M_\odot$ stars. Unlike canonical stars, a 2.01 $M_\odot$ can possess both hyperons and $\Delta$'s. Therefore, I analyze the reduction in the radii separately. The values of $\Delta R$ are presented in the order varying $L$ from 44 MeV up to 92 MeV as previously discussed in the text.

Starting with the L3$\omega\rho$ model, the NY matter present $\Delta R$ equals 0.23 km (1.77 $\%$), 0.25 km (2.08 $\%$), 0.31 km (2.46 $\%$), and 0.35 km (2.68 $\%$). In contrast, for the BigApple, the values of $\Delta R$ are: 0.12 km (0.09 $\%$), 0.07 km (0.5\%), 0.07 km (0.5\%), and 0.04 km (0.3\%). Thus, it is evident that a 2.01 $M_\odot$ in the L3$\omega\rho$ is more sensitive to changes in the slope than in the BigApple. However, more importantly, as the slope increases, $\Delta R$ values rise for the L3$\omega\rho$ but decrease for the BigApple.

In the case of ND matter, for the L3$\omega\rho$, the values of $\Delta R$ we obtain: 0.41 km (3.31 $\%$), 0.41 km (3.27 $\%$), 0.37 km (2.93 $\%$),  and 0.11 km (0.84 $\%$). On the other hand, for the BigApple we have: 0.32 km (2.4 $\%$), 0.35 km (2.59 $\%$), 0.33 km (2.42 $\%$),  and 0.24 km (1.96 $\%$). In the case of 2.01 $M_\odot$, qualitatively, we can notice that the effects of the $\Delta^-$ particle are reduced as we increase the slope for both parametrizations. Quantitatively, the effects here are more pronounced for the L3$\omega\rho$ than for the BigApple, as done for the case of NY matter.

The last possibility is  NYD matter. Due to the combined effect of both, $\Delta^-$ and hyperons, NYD matter presents the largest values of $\Delta R $. For the L3$\omega\rho$ we obtain by increasing $L$ from 44 MeV up to 92 MeV, $\Delta R = $ 0.63 km (5.08$\%$), 0.64 km (5.13 $\%$), 0.62 km (4.92 $\%$), and 0.42 km (3.22 $\%$). In the case of the BigApple, the values of $\Delta R$ are: 0.40 km (3.00\%),  0.42 km (3.11\%),  0.41 km (3.01\%), and 0.35 km (2.53 \%). As can be seen, in the case of the NYD matter, the effects of $L$ are exacerbated when compared with the NY and ND matter, although there is a clear reduction in $\Delta R$ for $L = 92$ MeV for both models.

All the relevant parameters are summarized in Tab.~\ref{T3}.

\subsection{Dimensionless tidal parameter}

%%%%%%%%%%%%%%%%%
\begin{figure*}[ht!]
\begin{tabular}{ccc}
\centering % \begin{center}/\end{center} takes some additional vertical space
\includegraphics[scale=.52, angle=270]{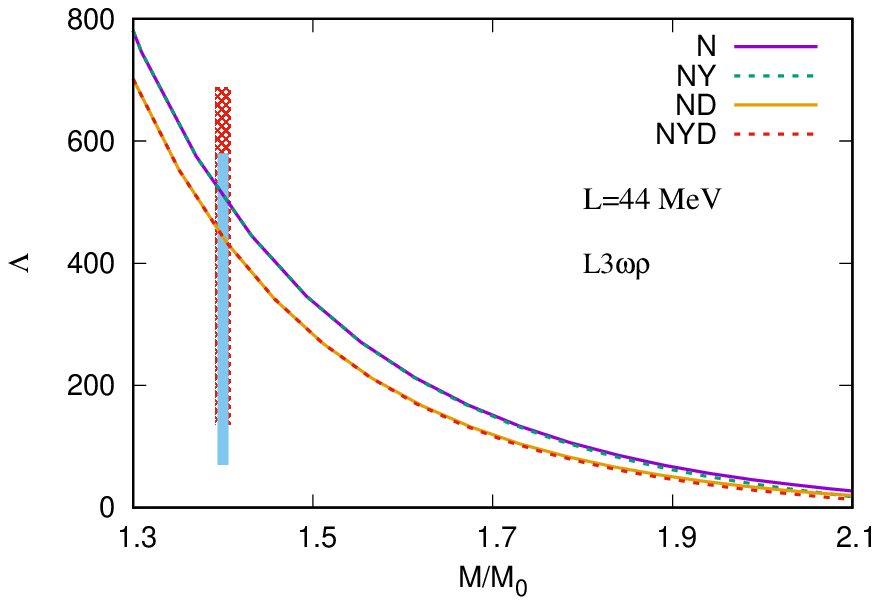} &
\includegraphics[scale=.52, angle=270]{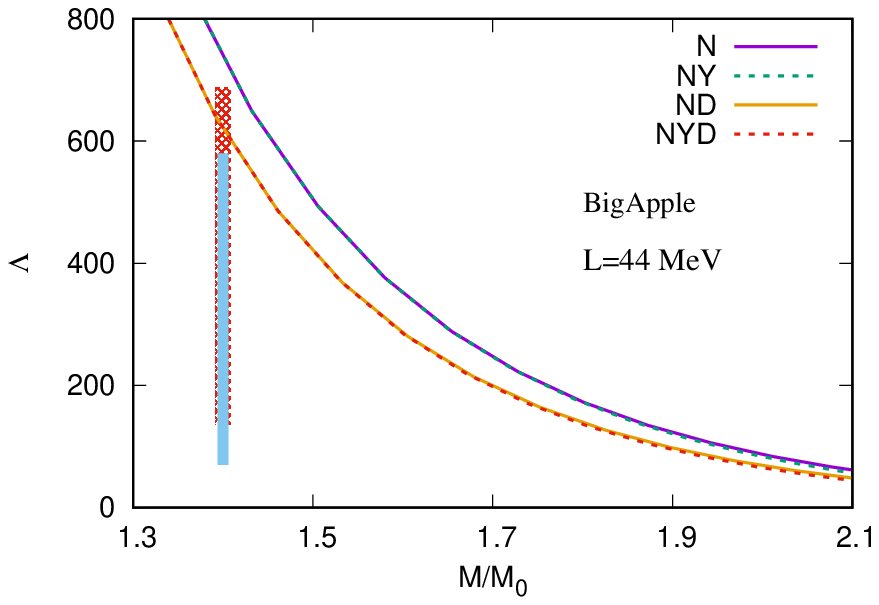} \\
\includegraphics[scale=.52, angle=270]{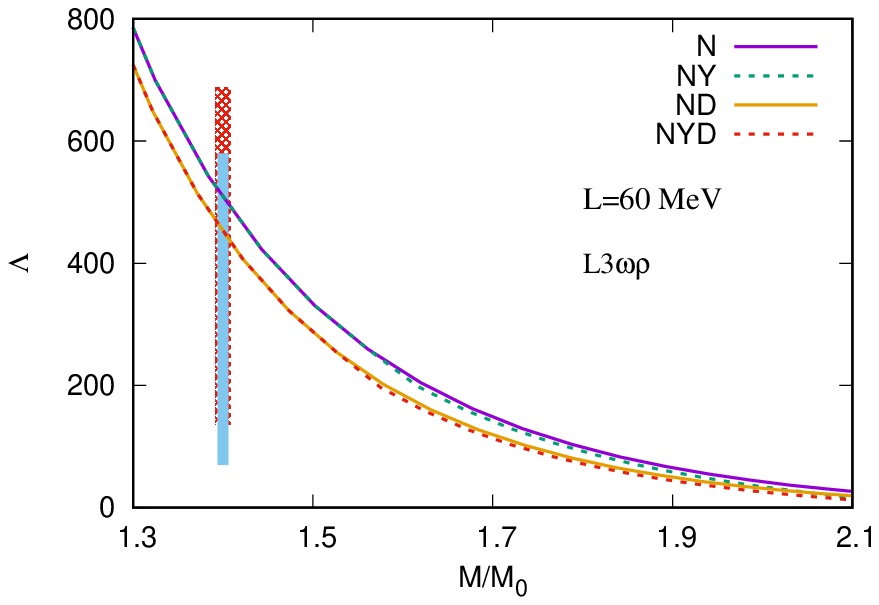} &
\includegraphics[scale=.52, angle=270]{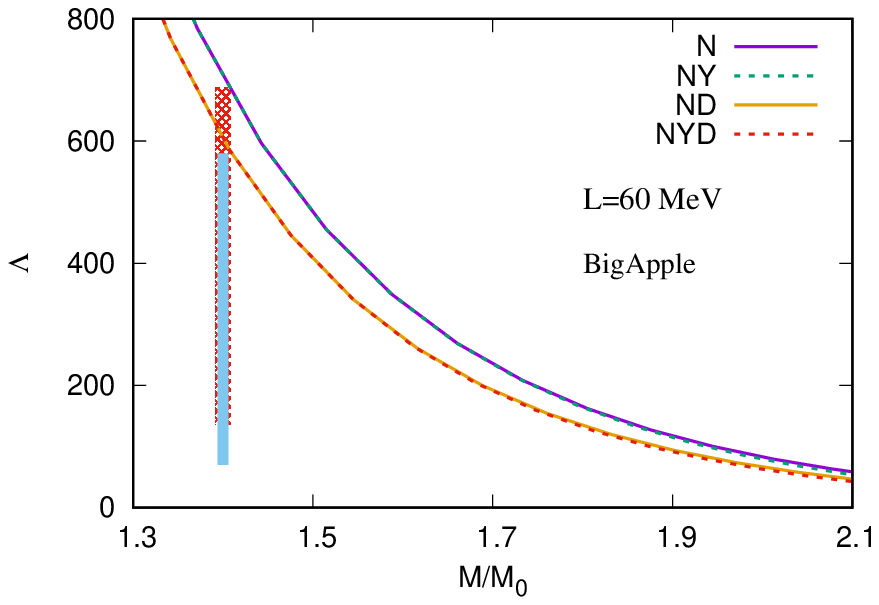}\\
\includegraphics[scale=.52, angle=270]{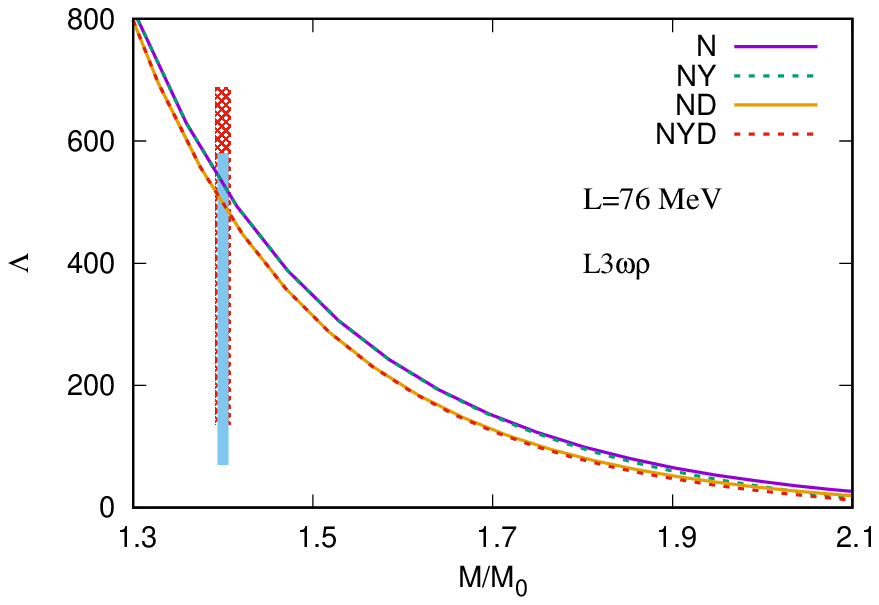} &
\includegraphics[scale=.52, angle=270]{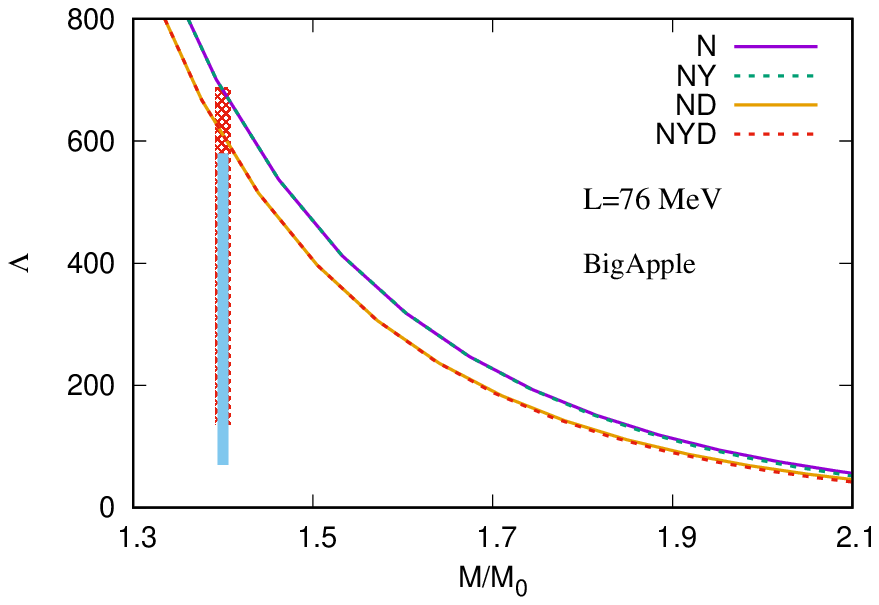}\\
\includegraphics[scale=.52, angle=270]{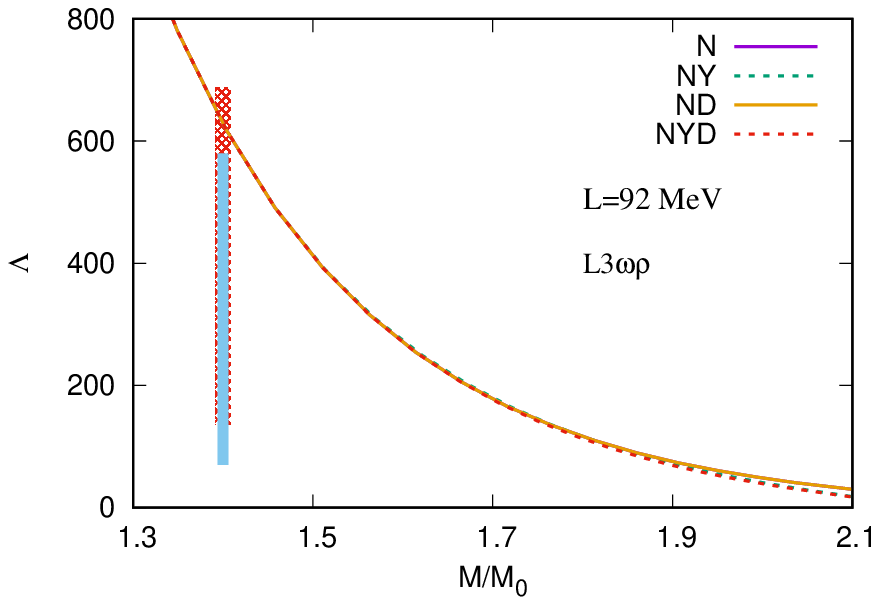} &
\includegraphics[scale=.52, angle=270]{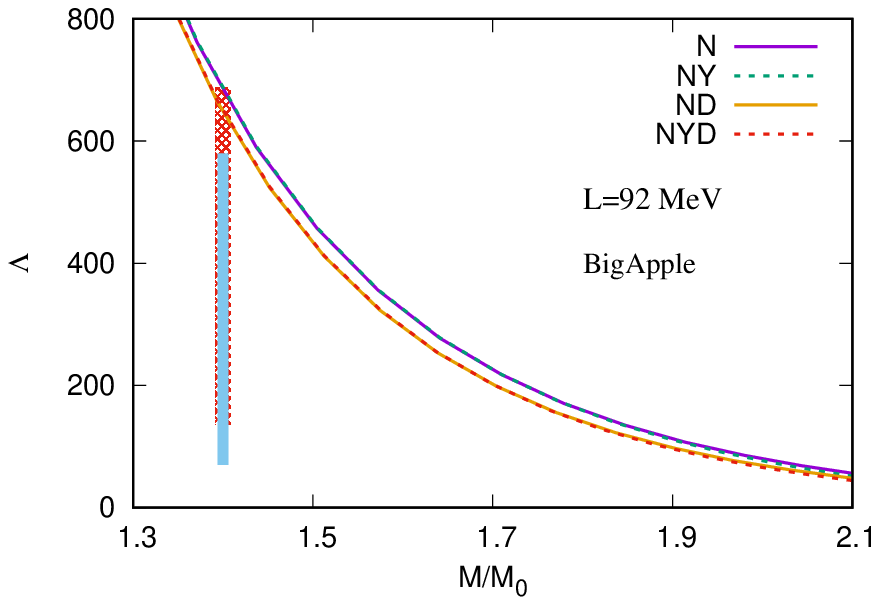}\\
\end{tabular}
\caption{Dimensionless tidal parameter $\Lambda$ for different values of $L$ and different neutron stars' interiors. The hatched areas are constraints as discussed in the text.} \label{tidaleps}
\end{figure*}

{In recent years, the theory of relativistic tidal effects in binary systems has been the focus of intense research. Below, I provide a summary of the procedure for computing the dimensionless tidal parameter $\Lambda$, which measures how easily an object is deformed by an external tidal field. It is defined as follows:
\begin{equation}
 \Lambda = \frac{2k_2}{3C^5} , \label{stidal}
\end{equation}
where $M$ is the mass of the compact object and $C = GM/R$ is its compactness. The parameter $k_2$ is known as the second-order Love number and is given by (see Ref. \cite{Hinderer_2008}):}
\begin{equation}
\begin{aligned} 
k_{2}= & \frac{8 C^{5}}{5}(1-2 C)^{2}\left[2-y_{R}+2 C\left(y_{R}-1\right)\right] \\ & \times\left\{2 C\left[6-3 y_{R}+3 C\left(5 y_{R}-8\right)\right]\right. \\ & +4 C^{3}\left[13-11 y_{R}+C\left(3 y_{R}-2\right)+2 C^{2}\left(1+y_{R}\right)\right] \\ & \left.+3(1-2 C)^{2}\left[2-y_{R}+2 C\left(y_{R}-1\right)\right] \ln (1-2 C)\right\}^{-1}
\end{aligned}
\end{equation}
where $y_R=y(r=R)$ and $y(r)$ is obtained by solving:
\begin{equation}
 r\frac{dy}{dr} +y^2 + yF(r) +r^2Q(r) = 0 . \label{EL15}
\end{equation}
{The coefficients $F(r)$ and $Q(r)$ are given by:
\begin{eqnarray} 
F(r) &= & {\left[1-4 \pi G r^{2}(\epsilon-p)\right]\left[1-\frac{2GM}{r}\right]^{-1} } , \\ 
Q(r) & = & 4 \pi G\left[5 \epsilon+9 p+\frac{\epsilon+p}{v_{s}^{2}}-\frac{6}{4 \pi G r^{2}}\right]\left[1-\frac{2 GM}{r}\right]^{-1}   \nonumber \\ 
& & -\frac{4 GM^{2}}{r^{4}}\left[1+\frac{4 \pi r^{3} p}{M}\right]^{2}\left[1-\frac{2 GM}{r}\right]^{-2} ,
\label{EL17}
\end{eqnarray}
where $v_{s}^{2} \equiv d p / d \epsilon$ is the squared speed of sound. The boundary condition for Eq. \eqref{EL15}  at $r = 0$ is given by $y(0) = 2$.
}

The dimensionless tidal parameter $\Lambda$ is displayed in Fig.~\ref{tidaleps} for the L3$\omega\rho$ and the BigApple. Altogether, I also display two constraints. As in the case of the radius of the canonical star, I choose a strong constraint and a more conservative one. The strong constraint presented in ref.~\cite{AbbottPRL} points to the canonical star $70~<~\Lambda_{1.4}~<~580$, coming directly from the analysis of the GW170817 event, and it is displayed as a bluish hatched area. The conservative constraint combines the data coming from GW170817 with Bayesian analyses and points to $133~<\Lambda_{1.4}~<686$ ~\cite{Li2021}, and it is presented as a reddish hatched area.  As in the case of the radius of the canonical star, both strong and weak constraints are required for the dimensionless tidal parameter to allow for a broad exploration of the slope parameter $L$. 

We can notice that the change in the symmetry energy slope causes opposite effects in the dimensionless tidal parameter for the two parametrizations. While for the L3$\omega\rho$, increasing $L$ also increases $\Lambda$, especially for the canonical star, $\Lambda_{1.4}$, varying from 515 to $L = 44$ MeV, up to 639 for $L$ = 92 MeV; although the minimum happens at $L = 60$ MeV. In the case of the BigApple, increasing $L$ actually reduces $\Lambda$, from 745 to 708, while the minimum occurs at $L = 76$ MeV.  Such a feature was already pointed out in ref.~\cite{lopes2024PRCb}.

In relation to the neutron stars' exotic content, we know that 1.4 $M_\odot$ stars do not present hyperons; therefore, I only analyze here the effects of $\Delta$ resonances. Despite the opposite effects $L$ for pure nucleonic stars, the effects of the slope for $\Delta$-admixed neutron stars are the same for the L3$\omega\rho$ and the BigApple: The presence of $\Delta's$ reduces the value of $\Lambda_{1.4}$ for a fixed slope, while at the same time, increasing the slope reduces the values of $\Delta \Lambda_{1.4}$, implying that stars with lower values of $L$ are more sensitive to the presence of exotic matter. Quantitatively, the reduction for the BigApple is more significant than the one in the L3$\omega\rho$, which is curious because the influence on the radius is the opposite.
In the case of the L3$\omega\rho$, from $L = 44$ MeV up to 92 MeV, we have $\Delta \Lambda_{1.4}$ varying from 64 (12.5\%) up to 0 due to the strong suppression of $\Delta$ particles, as shown in Fig.~\ref{DA} and Fig.~\ref{YD}. In relation to the BigApple, the values of $\Delta \Lambda_{1.4}$ decrease from 119 to 49 as the slope is increased.

Regarding the bounds related to the GW170817 event, it can be seen that the L3$\omega\rho$ can satisfy the strong constraint ($70~<~\Lambda_{1.4}~<580$) for all values of $L$ except $L = 92$ MeV, even in the absence of exotic content. For this value of the slope, only the conservative constraint $(\Lambda_{1.4}~<686)$ is achieved. As for $L = 92$ MeV, the amount of $\Delta$ particles is insignificant;  the strong constraint is never fulfilled for this value of slope, a direct consequence of having $\Lambda_{\omega\rho} = 0$ for this value of $L$. In opposition, not even the conservative constraint is achieved for pure nucleonic matter in the case of the BigApple. When $\Delta$ resonances are present, only the conservative constraint is fulfilled for all values of $L$.  This indicates that the BigApple is disfavored in relation to the L3$\omega\rho$. However, the confirmation of the existence of very massive stars, such as the PSR J0952-0607~\cite{romani}, combined with at least a moderate value of $\Lambda_{1.4}$, can potentially change this picture.

The  values of $\Lambda_{1.4}$ and $\Delta \Lambda_{1.4}$ are displayed in Tab.~\ref{T3}. 

\subsection{Radius Difference between a canonical 1.4$M_\odot$ and a 2.01$M_\odot$.}

Recently, it was suggested in ref.~\cite{Tang2025PRD} that the radius difference between a 1.4$M_\odot$ and a 2.0$M_\odot$ star is linked to the symmetry slope $L$. 
The authors reported a positive correlation: larger values of $L$ are associated with larger values of $R_{1.4} - R_{2.0}$. However, this difference is centered at a negative value, indicating that $R_{2.0}~>~R_{1.4}$.

 Here, I extend this study, analyzing not only how different models and different slopes affect this difference, but also investigating the role played by different exotic content in neutron stars' interiors. I began defining the quantity $\delta R$:
 \begin{equation}
\delta R = R_{1.4} - R_{2.01},
 \end{equation}
 where, only to keep the internal consistency, I use 2.01 $M_\odot$ instead of 2.0$M_\odot$. The results are presented in km in the last column of Tab.~\ref{T3}.

%\newpage

\begin{widetext}
\begin{center}
\begin{table}[ht!]
\begin{center}
\scriptsize
\begin{tabular}{ccc|ccccccccccc}
\toprule
Model &  $L$ (MeV) & Content & $R_{1.4}$~ & $R_{2.01}$ ~& $\Lambda_{1.4}$ & $M_{max}/M_\odot$ & $\Delta R_{1.4}$  & $\Delta R_{2.01}$ & $\Delta\Lambda_{1.4}$ & $\Lambda^0$ (fm$^{-3}$) & $\Xi^-$ (fm$^{-3}$) & $\Delta^-$ (fm$^{-3}$) & $\delta R$\\
\toprule
L3$\omega\rho$ & 44 & N & 12.58  & 12.40 & 515 & 2.31 & - & - &  - & - & - & - & +0.18 \\
 %\hline
L3$\omega\rho$ & 60 & N  & 12.74 & 12.47 & 513 & 2.30 & - & - & - & - & - & - & +0.27 \\
%  \hline
L3$\omega\rho$ & 76& N &  12.99& 12.59& 535 & 2.30 & - & - & - & - &- & - & +0.40        \\
% \hline
L3$\omega\rho$ &92 &  N  & 13.48 & 13.04 & 639 & 2.34 & - & - & - & - & - & - & +0.44  \\
\hline
L3$\omega\rho$ & 44 & NY &  12.58  & 12.17 & 515 & 2.14 & 0.0 & 1.9$\%$ & 0.0 & 0.437 & 0.537 & -  &  +0.41 \\
 %\hline
L3$\omega\rho$ & 60 & NY  & 12.74 & 12.22 & 513 & 2.13 & 0.0 & 2.1\% & 0.0 & 0.437 & 0.521 & - &  +0.52 \\
%  \hline
L3$\omega\rho$ & 76& NY &  12.99 & 12.28 & 535 & 2.12 & 0.0 & 2.5\% & 0.0 & 0.438 & 0.479 & - & +0.71     \\
% \hline

L3$\omega\rho$ &92 &  NY  & 13.48 & 12.69 & 639 & 2.13 & 0.0 & 2.7$\%$ & 0.0 & 0.475 & 0.408 & -  &  +0.79  \\
\hline
L3$\omega\rho$ & 44 & ND &  12.38 & 11.99 & 451 & 2.29 & 1.6$\%$ & 3.3$\%$ & 12.5$\%$ & - & - & 0.322 & +0.39  \\
 %\hline
L3$\omega\rho$ & 60 & ND  & 12.57 & 12.06 & 464 & 2.28 & 1.3\% & 3.3\% & 9.6\% & - & - & 0.326 & +0.51 \\
%  \hline
L3$\omega\rho$& 76&  ND &  12.89 & 12.22 & 503 & 2.28 & 0.8\% & 2.9$\%$ & 5.9\% & - & - & 0.337  & +0.67         \\
L3$\omega\rho$ &92 &  ND  &  13.48 & 12.93 & 639 & 2.33 & 0.0 & 0.8\% & 0.0 & - & - & 0.371 & +0.55   \\
\hline
L3$\omega\rho$& 44 & NYD & 12.38 & 11.77 & 451 & 2.14 & 1.6\% & 5.1\% & 12.5\% & 0.476 & 0.772 & 0.322 & +0.61 \\
 %\hline
L3$\omega\rho$ & 60 & NYD  & 12.57 &  11.83 & 464 & 2.13 & 1.3\% & 5.1\% & 9.6\% & 0.472 & 0.739 & 0.326 & +0.74 \\
%  \hline
L3$\omega\rho$ & 76& NYD &  12.89 & 11.97 & 503 & 2.12 & 0.8\% & 4.9\% & 5.9\% & 0.465 & 0.656 & 0.337 & +0.92        \\
% \hline
L3$\omega\rho$ &92 &  NYD  & 13.48 & 12.63 & 639 & 2.13 & 0.0 & 3.1\% & 0.0 & 0.472 & 0.422 & 0.371 & +0.85  \\
\hline
BigApple & 44 & N &  13.03  & 13.34 & 745  & 2.60 & - & - & - & - & - & - & -0.31 \\
 %\hline
BigApple & 60 & N  &  13.31 & 13.49 & 724  &2.60 & - & - & - & - & - & - & -0.18 \\
%  \hline
BigApple & 76& N & 13.48 & 13.59 &  699 & 2.60  & - & - & - & - & - & - & -0.11    \\
% \hline
BigApple &92 &  N & 13.77   &  13.80 & 708 & 2.60 & - & - & - & - & - & - & -0.03  \\
\hline
BigApple & 44 & NY &  13.03 & 13.22 & 745 & 2.41 & 0.0 & 0.9\% & 0.0 & 0.364 & 0.462 & - & -0.19\\
 %\hline
BigApple & 60 & NY  &  13.31 & 13.42 & 724  & 2.41 & 0.0 & 0.5\% & 0.0 & 0.364 & 0.449 & - & -0.11 \\
%  \hline
BigApple & 76& NY & 13.48 & 13.52 &  699 & 2.40 & 0.0 & 0.5\% & 0.0 & 0.364 & 0.428 & - & -0.04\\
% \hline
BigApple &92 &  NY & 13.77  &  13.76 & 708 & 2.40 &  0.0 & 0.3\% & 0.0 & 0.356 & 0.392 & - & +0.01  \\
\hline
BigApple & 44 & ND &  12.85 & 13.02 & 626 & 2.59 & 1.4\% & 2.4\% & 16.0\% & - & - & 0.268 & -0.17\\
 %\hline
BigApple & 60 & ND  &  13.07 & 13.14 & 609  & 2.59 & 1.8\% & 2.6\% & 15.8\% & - & - & 0.271 & -0.07 \\
%  \hline
BigApple & 76& ND & 13.28 & 13.26 &  624 & 2.59 & 1.5\% & 2.4\% & 10.7\% &- & - & 0.274 & +0.02    \\
% \hline
BigApple &92 &  ND & 13.64  &  13.53 & 659 & 2.59 &  1.0\% & 2.0\% & 6.9\% & - & - & 0.283 & +0.11\\
\hline
BigApple & 44 & NYD &  12.85 & 12.94 & 626 &2.42 &1.4\% & 3.0\% & 16.0\% & 0.378 & 0.626 & 0.268 & -0.11  \\
 %\hline
BigApple & 60 & NYD  &  13.07 & 13.07 & 609  & 2.42 & 1.8\% & 3.1\% & 15.8\% & 0.378 & 0.612 & 0.271 & 0.00 \\
%  \hline
BigApple & 76& NYD & 13.28 & 13.18 & 624  & 2.41 & 1.5\% & 3.0\% & 10.7\% & 0.378 & 0.577 & 0.274 & +0.10    \\
% \hline
BigApple &92 &  NYD & 13.64  &  13.45 & 659 &2.40 & 1.0\% & 2.5\% &6.9\% & 0.372 & 0.487   & 0.283 & +0.19  \\
\hline
\toprule 
\end{tabular}
\caption{ Some of the neutron stars' main properties and constraints. The radii are presented in km.} 
\label{T3}
\end{center}
\end{table}
\end{center}
\end{widetext}

The first feature we can notice is that when an exotic phase is absent, the L3$\omega\rho$ and the BigApple present opposite behavior. Although the positive correlation between $\delta R$ and $L$ is present in both, the values of $\delta R$ are always positive for the L3$\omega\rho$ and always negative for the BigApple. This implies that as we increase the slope, the 1.4$M_\odot$ and the 2.01 $M_\odot$ stars become more and more differentiable in the L3$\omega\rho$, as the absolute value of $\delta R$ increases; but become more and more similar for the BigApple as the absolute value of $\delta R$ decreases. 

When hyperons are present, massive neutron stars become even more compact. For the L3$\omega\rho$ model, this causes a value of $\delta R$ being as large as 0.8 km in the case of large slopes, while for the BigApple, the difference becomes negligible.
$\Delta$ resonances have a similar effect to hyperons, and $\Delta$-admixed hyperonic neutron stars have the larger positive correlation between $\delta R$ and $L$. Furthermore, the values of $\delta R$ can be as high as 0.92 km.

These behaviors indicate that a precise measure of the radii of massive and canonical stars can not only potentially indicate the presence of hyperons, but also can help us to choose which model has a better description of neutron stars. For instance, a $R_{2.01}~<$ 12 km can only be obtained with the L3$\omega\rho$ with an exotic phase. In the same sense, a $\delta R~<0$ may favor the BigApple parametrization, although the analysis of  $\delta R$ must be done altogether with the other constraints as discussed in the text.

A somewhat related quantity was studied in Refs.~\cite{Marcio2025PRD,BausweinPRR2026}, the slope of the neutron star mass-radius curve, $dM/dR$. In Ref.~\cite{Marcio2025PRD}, the authors compare several equations of state and show that only 1\% of hyperonic EOSs that satisfy some observational constraints have a negative slope for the canonical 1.4$M_\odot$, while Ref.~\cite{BausweinPRR2026} studies the second derivative of the mass-radius curve. Considering only N and NY matter, in the present study, only the L3$\omega\rho$ parametrization with $L =76$ MeV and $L = 92$ MeV has a negative slope of the mass-radius curve for the canonical star. If hyperons are indeed inevitable, this feature favors lower values of $L$.

\section{Conclusions}

In this work, I studied the effects of the slope $L$ in the neutron stars' exotic content.
The main results are summarized below.

\begin{itemize}

    \item For the NY matter, increasing the slope does not significantly affect the onset of $\Lambda^0$ hyperons, but suppresses them at high densities. The effects on the $\Xi^-$ are the opposite. Increasing $L$ increases their population.

    \item For ND matter, increasing $L$ also reduces the presence of $\Delta^-$ resonances at high densities. But unlike the $\Lambda^0$ particle, which has a monotonically increasing behavior,  the fraction of  $\Delta^-$ began to drop at high densities. For the NYD matter, the drop of the population of $\Delta^-$ resonances is even more evident. The exarcebed behaviour found for $L = 92$ MeV within the L$3\omega\rho$ is not exclusively due to the large slope, but also because such a parametrization possesses $\Lambda_{\omega\rho} = 0$.

    \item The slope affects the  square of the speed of sound, $v_s^2$. At low densities, lower values of slope have a small value of $v_s^2$, while large values of $L$ present a large value of $v_s^2$.     
    For nucleonic stars, there is a peak in the curve for the BigApple parametrization, but not for the L3$\omega\rho$, reflecting the quartic term for the $\omega$ field $\xi$.

    \item Qualitatively, the effects of the exotic phase on $v_s^2$ are the same for the L3$\omega\rho$ and the BigApple, although $v_s^2$ presents a steep growth for the BigApple.

    \item The onset of hyperons causes a first drop in $v_s^2$ followed by a small second drop due to the onset of the second hyperon.

    \item The presence of $\Delta's$ causes a large drop followed by a steep increase in $v_s^2$. $L = 92$ MeV for the L3$\omega\rho$ present a qualitate different behaviour because it also have $\Lambda_{\omega\rho} = 0$, which allow large values for the $\rho$ field.

    \item Increasing $L$ also increases the stars' radii, a well-known result. Hyperons do not affect the canonical $1.40M_\odot$, but a $2.01M_\odot$ star has its radius reduced due to the onset of hyperons. The lower the slope, the larger the reduction. The variations $\Delta R_{2.01}$ are larger for the L3$\omega\rho$.

    \item Unlike hyperons, $\Delta$ resonances affect both the canonical 1.4$M_\odot$ and a  2.01$M_\odot$. In general, the lower the slope, the larger the value of $\Delta R_{1.4}$ and $\Delta R_{2.01}$.

    \item $\Delta's$ also affects $\Lambda_{1.4}$. More than that, the effects are even larger than the effects in the radius, reaching 16$\%$ for the BigApple.  Again, the effects are larger for lower values of $L$.

    \item { In relation to the constraints related to the dimensionless tidal parameter, the L3$\omega\rho$ and the BigApple exhibit opposite behaviour. While the L3$\omega\rho$ can fulfill even the strong constraint for almost all values of $L$, the BigApple cannot describe even the weak bound, unless $\Delta$ particles are present.}

    \item A precise determination of both $R_{1.4}$ and $R_{2.01}$ will imply in a precise determination of $\delta R$. This quantity { when combined with additional constraint from astrophysical observation} can potentially not only indicate the presence of exotic matter in the core of neutron stars but also suggest which parametrization is better suited to describe these objects.
    
\end{itemize}

\textbf{Acknowledgments:}L.L.L.  was partially supported by CNPq (Brazil)
under Grant No 305347/2024-1.

\appendix

\counterwithin{figure}{section}

$$ $$

%%%%%%%%%%%%%%%%%%%%%%%%%
\bibliography{aref}

@article{lopescesar,
    title = {Imprints of the nuclear symmetry energy slope in gravitational wave signals emanating from neutron stars},
  author = {Lopes, L. L. and others},
  journal = {Phys. Rev. D},
  volume = {108},
  issue = {8},
  pages = {083042},
  numpages = {11},
  year = {2023},
  month = {Oct},
  publisher = {American Physical Society},
  doi = {10.1103/PhysRevD.108.083042},
  url = {https://link.aps.org/doi/10.1103/PhysRevD.108.083042}
}

@book{Glenbook,
  title={Compact stars:},
  author={Glendenning, Norman K},
  year={2000},
  publisher={2 ed. Edition, Springer New York}
}

@article{Serot_1992,
	doi = {10.1088/0034-4885/55/11/001},
	url = {https://doi.org/10.1088/0034-4885/55/11/001},
	year = 1992,
	month = {nov},
	publisher = {{IOP} Publishing},
	volume = {55},
	number = {11},
	pages = {1855--1946},
	author = {B D Serot},
	title = {Quantum hadrodynamics},
	journal = {Rep.  Progr.  Phys.}
}

@article{klahn2006,
  title = {Constraints on the high-density nuclear equation of state from the phenomenology of compact stars and heavy-ion collisions},
  author = {Kl\"ahn, T. and others},
  journal = {Phys. Rev. C},
  volume = {74},
  issue = {3},
  pages = {035802},
  numpages = {15},
  year = {2006},
  month = {Sep},
  publisher = {American Physical Society},
  doi = {10.1103/PhysRevC.74.035802},
  url = {https://link.aps.org/doi/10.1103/PhysRevC.74.035802}
}

@article{Riley:2019yda,
    author = "Riley, Thomas E. and others",
    title = "{A NICER View of PSR J0030+0451: Millisecond Pulsar Parameter Estimation}",
    doi = "10.3847/2041-8213/ab481c",
    journal = "Astrophys. J. Lett.",
    volume = "887",
    number = "1",
    pages = "L21",
    year = "2019"
}

@article{Miller:2019cac,
    author = "Miller, M.C. and others",
    title = "{PSR J0030+0451 Mass and Radius from $NICER$ Data and Implications for the Properties of Neutron Star Matter}",
    doi = "10.3847/2041-8213/ab50c5",
    journal = "Astrophys. J. Lett.",
    volume = "887",
    number = "1",
    pages = "L24",
    year = "2019"
}

@article{Riley2021,
doi = {10.3847/2041-8213/ac0a81},
url = {https://dx.doi.org/10.3847/2041-8213/ac0a81},
year = {2021},
month = {sep},
publisher = {The American Astronomical Society},
volume = {918},
number = {2},
pages = {L27},
author = {Riley, T.E. and others},
title = "{A NICER View of the Massive Pulsar PSR J0740+6620 Informed by Radio Timing and XMM-Newton Spectroscopy}",
journal = {Astrophys. J. Lett.}
}

@article{Annala2018PRL,
  title = {Gravitational-Wave Constraints on the Neutron-Star-Matter Equation of State},
  author = {Annala, Eemeli and Gorda, Tyler and Kurkela, Aleksi and Vuorinen, Aleksi},
  journal = {Phys. Rev. Lett.},
  volume = {120},
  issue = {17},
  pages = {172703},
  numpages = {5},
  year = {2018},
  doi = {10.1103/PhysRevLett.120.172703},
  url = {https://link.aps.org/doi/10.1103/PhysRevLett.120.172703}
}

@article{lopes2023ptep,
    author = {Lopes, Luiz L},
    title = "{A closer look at the Yukawa interaction from a symmetry group perspective}",
    journal = {Prog.  Theor.  Exper. Phys.},
    volume = {2023},
    number = {11},
    pages = {113D01},
    year = {2023},
    issn = {2050-3911},
    doi = {10.1093/ptep/ptad129},
    }

@article{Miller2021,
	doi = {10.3847/2041-8213/ac089b},
	url = {https://doi.org/10.3847/2041-8213/ac089b},
	year = 2021,
	month = {sep},
	publisher = {American Astronomical Society},
	volume = {918},
	number = {2},
	pages = {L28},
	author = {Miller, M.C. and others},
	title = "{The Radius of PSR J0740+6620 from NICER and XMM-Newton Data}",
	journal = {Astrophys. J. Lett.}
}

@ARTICLE{Paar2014,
   title = {Neutron star structure and collective excitations of finite nuclei},
  author = {Paar, N. and others},
  journal = {Phys. Rev. C},
  volume = {90},
  issue = {1},
  pages = {011304},
  numpages = {4},
  year = {2014},
  month = {Jul},
  publisher = {American Physical Society},
  doi = {10.1103/PhysRevC.90.011304},
  url = {https://link.aps.org/doi/10.1103/PhysRevC.90.011304}
}

@ARTICLE{Lattimer2013,
   title = {CONSTRAINING THE SYMMETRY PARAMETERS OF THE NUCLEAR INTERACTION},
   author = {J. Lattimer and Y. Lim},
          journal = {Astrophys. J.},
         year = 2013,
       volume = {771},
        pages = {51},
          doi = { 10.1088/0004-637X/771/1/51},
}

@ARTICLE{Steiner2014,
       author = {J. Lattimer and A. Steiner},
       title = {Constraints on the symmetry energy using the mass-radius relation of neutron stars},
          journal = {Eur. Phys. J. A},
         year = 2014,
       volume = {50},
        pages = {40},
          doi = {10.1140/epja/i2014-14040-y},
}

@ARTICLE{pions,
       author = {J. Estee and others},
       title = {Probing the Symmetry Energy with the Spectral Pion Ratio},
        journal = {Phys. Rev. Lett.},
         year = 2020,
       volume = {126},
        pages = {162701},
          doi = {10.1103/PhysRevLett.126.162701},
          }

@ARTICLE{PREX2,
       author = {B. Reed and others},
       title = {Implications of PREX-2 on the Equation of State of Neutron-Rich Matter},
       journal = {Phys. Rev. Lett.},
         year = 2020,
       volume = {126},
        pages = {172503},
          doi = {10.1103/PhysRevLett.126.172503},
          }

@article{romani,
doi = {10.3847/2041-8213/ac8007},
url = {https://dx.doi.org/10.3847/2041-8213/ac8007},
year = {2022},
month = {jul},
publisher = {The American Astronomical Society},
volume = {934},
number = {2},
pages = {L17},
author = {Roger W. Romani and D. Kandel and Alexei V. Filippenko and Thomas G. Brink and WeiKang Zheng},
title = {PSR J0952-0607: The Fastest and Heaviest Known Galactic Neutron Star},
journal = {Astrophys. J. Lett.},
}

@article{TOV,
  title = {On Massive Neutron Cores},
  author = {Oppenheimer, J. R. and Volkoff, G. M.},
  journal = {Phys. Rev.},
  volume = {55},
  issue = {4},
  pages = {374--381},
  numpages = {0},
  year = {1939},
  month = {Feb},
  publisher = {American Physical Society},
  doi = {10.1103/PhysRev.55.374},
  url = {https://link.aps.org/doi/10.1103/PhysRev.55.374}
}

@article{Bodek2020,
  title = {Comparison of optical potential for nucleons and $\Delta$ resonances.},
  author = {A. Bodek and T. Cai},
  journal = {Eur. Phys. J. C},
  volume = {80},
  pages = {655},
  year = {2020},
  doi = {10.1140/epjc/s10052-020-8236-8},
  url = {https://doi.org/10.1140/epjc/s10052-020-8236-8}
}

@article{Li2021,
  title = {Constraining the nuclear symmetry energy and properties of the neutron star from GW170817 by Bayesian analysis},
  author = {Y. Li and H. Chen and D. Wen and others},
  journal = {Eur. Phys. J. A},
  volume = {57},
  pages = {31},
  year = {2021},
  doi = {10.1140/epja/s10050-021-00342-w},
  url = {https://doi.org/10.1140/epja/s10050-021-00342-w}
}

@article{Tang2025PRD,
  title = {Phase transition and nuclear symmetry energy from neutron star observations: Constraints in light of PSR J0614-3329},
  author = {Tang, Shao-Peng and Huang, Yong-Jia and Fan, Yi-Zhong},
  journal = {Phys. Rev. D},
  volume = {112},
  pages = {083009},
  year = {2025},
  publisher = {American Physical Society},
  doi = {10.1103/bmsk-8n85},
  url = {https://link.aps.org/doi/10.1103/bmsk-8n85}
}

@article{LopesUNIVERSE2025,
AUTHOR = {Lopes, Luiz L.},
TITLE = {An Undergraduate Approach to the Quantum Hadrodynamics and Physics of Neutron Stars},
JOURNAL = {Universe},
VOLUME = {11},
YEAR = {2025},
NUMBER = {8},
pages = {276},
URL = {https://www.mdpi.com/2218-1997/11/8/276},
ISSN = {2218-1997},
DOI = {10.3390/universe11080276}
}

@article{lopes2024PRCb,
  title = {Role of the symmetry energy slope in neutron stars: Exploring the model dependency},
  author = {Lopes, Luiz L.},
  journal = {Phys. Rev. C},
  volume = {110},
  pages = {015805},
  numpages = {11},
  year = {2024},
  doi = {10.1103/PhysRevC.110.015805},
  url = {https://link.aps.org/doi/10.1103/PhysRevC.110.015805}
}

@article{FriedmanEPJWC2022,
	author = {{Friedman, Eliahu} and {Gal, Avraham}},
	title = {$\Xi$-nuclear constraints from $\Xi-$ emulsion capture events},
	DOI= "10.1051/epjconf/202227103002",
	url= "https://doi.org/10.1051/epjconf/202227103002",
	journal = {EPJ Web Conf.},
	year = 2022,
	volume = 271,
	pages = "03002",
}

@article{FRIEDMAN2025PLB,
title = {Compatibility of recent $\Xi-$ nuclear bound state signals},
journal = {Phys. Lett. B},
volume = {868},
pages = {139728},
year = {2025},
issn = {0370-2693},
doi = {https://doi.org/10.1016/j.physletb.2025.139728},
url = {https://www.sciencedirect.com/science/article/pii/S0370269325004897},
author = {E. Friedman and A. Gal},
}

@article{HaidenbauerEPJA2023,
	author = {{Haidenbauer, Johann} and {MeiBner, Ulf-G.} and {Nogga, Andreas} and {Le, Hoai}},
	title = {Hyperon–nucleon interaction in chiral effective field theory at next-to-next-to-leading order},
	DOI= "10.1140/epja/s10050-023-00960-6",
	url= "https://doi.org/10.1140/epja/s10050-023-00960-6",
	journal = {Eur. Phys. J. A},
	year = 2023,
	volume = 59,
	number = 3,
	pages = "63",
}

@article{Marcio2025PRD,
  title = {Learning about neutron star composition from the slope of the mass-radius diagram},
  author = {Ferreira, M\'arcio and Provid\^encia, Constan\ifmmode \mbox{\c{c}}\else \c{c}\fi{}a},
  journal = {Phys. Rev. D},
  volume = {112},
  issue = {8},
  pages = {083058},
  numpages = {10},
  year = {2025},
  publisher = {American Physical Society},
  doi = {10.1103/r7gk-kcmn},
  url = {https://link.aps.org/doi/10.1103/r7gk-kcmn}
}

@article{BausweinPRR2026,
  title = {Stellar properties indicating the presence of hyperons in neutron stars},
  author = {Bauswein, Andreas and Nikolaidis, Aristeidis and Lioutas, Georgios and Kochankovski, Hristijan and Char, Prasanta and Mondal, Chiranjib and Oertel, Micaela and Tolos, Laura and Chamel, Nicolas and Goriely, Stephane},
  journal = {Phys. Rev. Res.},
  volume = {8},
  issue = {1},
  pages = {013253},
  numpages = {17},
  year = {2026},
  publisher = {American Physical Society},
  doi = {10.1103/ygtr-ktqk},
  url = {https://link.aps.org/doi/10.1103/ygtr-ktqk}
}

@article{Mauviard2025APJ,
doi = {10.3847/1538-4357/ae145d},
url = {https://doi.org/10.3847/1538-4357/ae145d},
year = {2025},
publisher = {The American Astronomical Society},
volume = {995},
number = {1},
pages = {60},
author = {Mauviard, Lucien and others},
title = {A NICER View of the 1.4 $M_0$ Edge-on Pulsar PSR J0614-3329},
journal = {Astrophys. J.},
}

@article{Choudhury2024APJL,
doi = {10.3847/2041-8213/ad5a6f},
url = {https://doi.org/10.3847/2041-8213/ad5a6f},
year = {2024},
publisher = {The American Astronomical Society},
volume = {971},
pages = {L20},
author = {Choudhury, Devarshi and others},
title = {A NICER View of the Nearest and Brightest Millisecond Pulsar: PSR J0437-4715},
journal = {Astrophys. J. Lett.},
}

@article{Mohammad2026APJ,
doi = {10.3847/1538-4357/ae18d0},
url = {https://doi.org/10.3847/1538-4357/ae18d0},
year = {2026},
publisher = {The American Astronomical Society},
volume = {997},
number = {1},
pages = {26},
author = {Mohammad Ali Looee, A. and Shahrbaf, M. and Moshfegh, H. R.},
title = {Hyperons in Neutron Stars across the Observed Mass Range: Insights from $\Lambda$N and $\Lambda\Lambda$ Interactions within a Microscopic Framework},
journal = {Astrophys. J.},
}

@article{Tong2025APJ,
doi = {10.3847/1538-4357/adba47},
url = {https://doi.org/10.3847/1538-4357/adba47},
year = {2025},
publisher = {The American Astronomical Society},
volume = {982},
number = {2},
pages = {164},
author = {Tong, Hui and Elhatisari, Serdar and MeiBner, Ulf-G.},
title = {Hyperneutron Stars from an Ab Initio Calculation},
journal = {Astrophys. J.},
}

@article{Potentials2000,
  title = {Properties of strange hadronic matter in bulk and in finite systems},
  author = {Schaffner-Bielich, J\"urgen and Gal, Avraham},
  journal = {Phys. Rev. C},
  volume = {62},
  issue = {3},
  pages = {034311},
  numpages = {8},
  year = {2000},
  month = {Aug},
  publisher = {American Physical Society},
  doi = {10.1103/PhysRevC.62.034311},
  url = {https://link.aps.org/doi/10.1103/PhysRevC.62.034311}}

@ARTICLE{LQCD,
   title = {Hyperon Forces from QCD and Their Applications},
        author = {T. Inoue},
          journal = {JPS Conf. Proc},
         year = 2019,
         volume = {26},
         pages = {023018},
       doi = {10.7566/JPSCP.26.023018},
       }

@article{Oliveira2007,
author = {de Oliveira, J. C. T. and Duarte, S. B. and Rodrigues, H. and Chiapparini, M. and Kyotoku, M.},
title = {EFFECTS OF $\Delta$-BARYON INTERACTION STRENGTH ON NEUTRON STARS PROPERTIES},
journal = {Int. J.  Mod. Phys. D},
volume = {16},
number = {02n03},
pages = {175-183},
year = {2007},
doi = {10.1142/S0218271807009929},
}

@article{KOLO2017,
title = {Delta isobars in relativistic mean-field models with $\sigma$-scaled hadron masses and couplings},
journal = {Nucl. Phys. A},
volume = {961},
pages = {106-141},
year = {2017},
issn = {0375-9474},
doi = {https://doi.org/10.1016/j.nuclphysa.2017.02.004},
url = {https://www.sciencedirect.com/science/article/pii/S0375947417300295},
author = {E.E. Kolomeitsev and K.A. Maslov and D.N. Voskresensky},
}

@article{Kauan2022PRC,
  title = {$\mathrm{\ensuremath{\Delta}}$ baryons in neutron stars},
  author = {Marquez, Kauan D. and Menezes, D\'ebora P. and Pais, Helena and Provid\^encia, Constan\ifmmode \mbox{\c{c}}\else \c{c}\fi{}a},
  journal = {Phys. Rev. C},
  volume = {106},
  issue = {5},
  pages = {055801},
  numpages = {14},
  year = {2022},
  month = {Nov},
  publisher = {American Physical Society},
  doi = {10.1103/PhysRevC.106.055801},
  url = {https://link.aps.org/doi/10.1103/PhysRevC.106.055801}
}

@article{Tagami2022,
title = {Slope parameters determined from CREX and PREX2},
journal = {Resul. Phys.},
volume = {43},
pages = {106037},
year = {2022},
issn = {2211-3797},
doi = {https://doi.org/10.1016/j.rinp.2022.106037},
url = {https://www.sciencedirect.com/science/article/pii/S2211379722006519},
author = {Shingo Tagami and Tomotsugu Wakasa and Masanobu Yahiro},
}

@article{AbbottPRL,
  title = {GW170817: Measurements of Neutron Star Radii and Equation of State},
  author = {Abbott, B. P. and Abbott, R. and Abbott, T. D. and others},
  journal = {Phys. Rev. Lett.},
  volume = {121},
  issue = {16},
  pages = {161101},
  numpages = {16},
  year = {2018},
  publisher = {American Physical Society},
  doi = {10.1103/PhysRevLett.121.161101},
  url = {https://link.aps.org/doi/10.1103/PhysRevLett.121.161101}
}

@ARTICLE{Abbott2017,
       author = {B. Abbott and others},
        title = {GW170817: Observation of Gravitational Waves from a Binary Neutron Star Inspiral},
          journal = {Phys. Rev. Lett.},
         year = 2017,
       volume = {119},
        pages = {161101},
          doi = {10.1103/PhysRevLett.119.161101},
}

@ARTICLE{IUFSU,
    title = {Relativistic effective interaction for nuclei, giant resonances, and neutron stars},
       author = {F. Fattoyev and others},
          journal = {Phys. Rev. C},
         year = 2010,
       volume = {82},
        pages = {055803},
          doi = {10.1103/PhysRevC.82.055803},
          }

@ARTICLE{lopes2024PRC,
 title = {Correlation between the symmetry energy slope and the deconfinement phase transition},
  author = {Lopes, Luiz L. and Menezes, Debora P. and Pelicer, Mateus R.},
  journal = {Phys. Rev. C},
  volume = {109},
  issue = {4},
  pages = {045801},
  numpages = {9},
  year = {2024},
  month = {Apr},
  publisher = {American Physical Society},
  doi = {10.1103/PhysRevC.109.045801},
  url = {https://link.aps.org/doi/10.1103/PhysRevC.109.045801}
}

@ARTICLE{Boguta,
    title = {Relativistic calculation of nuclear matter and the nuclear surface},
    author = {J. Boguta and  A.  Bodmer},
          journal = {Nucl. Phys. A},
         year = 1977,
       volume = {292},
        pages = {413},
          doi = {10.1016/0375-9474(77)90626-1},
          }

@ARTICLE{Rafa2011,
    title = {Neutron star properties and the symmetry energy
},

       author = {R.~Cavagnoli and D. Menezes and C.~Providencias},
          journal = {Phys. Rev. C},
         year = 2011,
       volume = {84},
        pages = {065810},
          doi = {10.1103/PhysRevC.84.065810},
          }

@ARTICLE{dex19jpg,
    title = {What do we learn about vector interactions from GW170817?},
       author = {V. Dexheimer and others},
          journal = {J. Phys. G},
         year = 2019,
       volume = {46},
        pages = {034002},
          doi = {10.1088/1361-6471/ab01f0},
          }

@article{KUBIS1997,
title = {Nuclear matter in relativistic mean field theory with isovector scalar meson},
journal = {Phys. Lett. B},
volume = {399},
number = {3},
pages = {191-195},
year = {1997},
issn = {0370-2693},
doi = {https://doi.org/10.1016/S0370-2693(97)00306-7},
url = {https://www.sciencedirect.com/science/article/pii/S0370269397003067},
author = {S. Kubis and M. Kutschera},
}

@article{Liu2002,
  title = {Asymmetric nuclear matter: The role of the isovector scalar channel},
  author = {Liu, B. and Greco, V. and Baran, V. and Colonna, M. and Di Toro, M.},
  journal = {Phys. Rev. C},
  volume = {65},
  issue = {4},
  pages = {045201},
  numpages = {11},
  year = {2002},
  month = {Mar},
  publisher = {American Physical Society},
  doi = {10.1103/PhysRevC.65.045201},
  url = {https://link.aps.org/doi/10.1103/PhysRevC.65.045201}
}

@article{Lopes2014BJP,
title = {Effects of the Symmetry Energy and its Slope on Neutron Star Properties},
journal = {Braz. J.  Phys.},
volume = {44},
pages = {774},
year = {2014},
doi = {10.1007/s13538-014-0252-4},
url = {https://link.springer.com/article/10.1007/s13538-014-0252-4},
author = {L.L. Lopes and D.P. Menezes},
}

@article{Lopes2022ApJ,
  title = {On the Nature of the Mass-gap Object in the GW190814 Event},
  author = {Luiz L. Lopes and Debora P. Menezes},
  journal = {Astrophys. J.},
  volume = {936},
  pages = {41},
  year = {2022},
  publisher = {American Astronomical Society},
  doi = {10.3847/1538-4357/ac81c4}, 
  }

@article{Lopes2024ApJ,
  title = {Decoding Rotating Neutron Stars: Role of the Symmetry Energy Slope},
  author = {Luiz L. Lopes},
  journal = {Astrophys. J.},
  volume = {966},
  pages = {184},
  year = {2024},
  publisher = {American Astronomical Society},
  doi = {10.3847/1538-4357/ad391e}, 
  }

@article{Lopes2022CTP,
doi = {10.1088/1572-9494/ac2297},
url = {https://dx.doi.org/10.1088/1572-9494/ac2297},
year = {2022},
month = {dec},
publisher = {IOP Publishing},
volume = {74},
number = {1},
pages = {015302},
author = {Luiz L Lopes},
title = {Hyperonic neutron stars: reconciliation between nuclear properties and NICER and LIGO/VIRGO results},
journal = {Commun.  Theor. Phys.},
}

@article{Dutra2014,
  title = {Relativistic mean-field hadronic models under nuclear matter constraints},
  author = {Dutra, M. and Louren\ifmmode \mbox{\c{c}}\else \c{c}\fi{}o, O. and Avancini, S. S. and others},
  journal = {Phys. Rev. C},
  volume = {90},
  issue = {5},
  pages = {055203},
  numpages = {35},
  year = {2014},
  month = {Nov},
  publisher = {American Physical Society},
  doi = {10.1103/PhysRevC.90.055203},
  url = {https://link.aps.org/doi/10.1103/PhysRevC.90.055203}
}

@article{Micaela2017,
  title = {Equations of state for supernovae and compact stars},
  author = {Oertel, M. and Hempel, M. and Kl\"ahn, T. and Typel, S.},
  journal = {Rev. Mod. Phys.},
  volume = {89},
  issue = {1},
  pages = {015007},
  numpages = {68},
  year = {2017},
  month = {Mar},
  publisher = {American Physical Society},
  doi = {10.1103/RevModPhys.89.015007},
  url = {https://link.aps.org/doi/10.1103/RevModPhys.89.015007}
}

@article{BPS,
  title = {THE GROUND STATE OF MATTER AT HIGH DENSITIES},
  author = {Gordon Baym and Christopher Pethick  and Peter Sutherland},
  journal = {Astrophys. J.},
  volume = {170},
  pages = {299},
  year = {1971},
  publisher = {American Astronomical Society},
  doi = {10.1086/151216}, 
  }

@article{BBP,
title = {Neutron star matter},
journal = {Nucl.Phys. A},
volume = {175},
number = {2},
pages = {225},
year = {1971},
issn = {0375-9474},
doi = {https://doi.org/10.1016/0375-9474(71)90281-8},
url = {https://www.sciencedirect.com/science/article/pii/0375947471902818},
author = {Gordon Baym and Hans A. Bethe and Christopher J Pethick},
}

@article{Hinderer_2008,
doi = {10.1086/533487},
url = {https://dx.doi.org/10.1086/533487},
year = {2008},
month = {apr},
publisher = {},
volume = {677},
number = {2},
pages = {1216},
author = {Tanja Hinderer},
title = {Tidal Love Numbers of Neutron Stars},
journal = {Astrophys. J.},
}

@article{Dapo2010,
  title={Appearance of hyperons in neutron stars},
  author={Dapo, H and Schaefer, B-J and Wambach, Jochen},
 journal = {Phys. Rev. C},
  volume = {81},
  issue = {3},
  pages = {035803},
  numpages = {11},
  year = {2010},
  month = {Mar},
  publisher = {American Physical Society},
  doi = {10.1103/PhysRevC.81.035803},
  url = {https://link.aps.org/doi/10.1103/PhysRevC.81.035803}
}

@article{BigApple,
  title = {GW190814: Impact of a 2.6 solar mass neutron star on the nucleonic equations of state},
  author = {Fattoyev, F. J. and Horowitz, C. J. and Piekarewicz, J. and Reed, Brendan},
  journal = {Phys. Rev. C},
  volume = {102},
  issue = {6},
  pages = {065805},
  numpages = {6},
  year = {2020},
  month = {Dec},
  publisher = {American Physical Society},
  doi = {10.1103/PhysRevC.102.065805},
  url = {https://link.aps.org/doi/10.1103/PhysRevC.102.065805}
}

@article{Miyatsu2013,
  title = {Equation of state for neutron stars in SU(3) flavor symmetry},
  author = {Miyatsu, Tsuyoshi and Cheoun, Myung-Ki and Saito, Koichi},
  journal = {Phys. Rev. C},
  volume = {88},
  issue = {1},
  pages = {015802},
  numpages = {14},
  year = {2013},
  doi = {10.1103/PhysRevC.88.015802},
  url = {https://link.aps.org/doi/10.1103/PhysRevC.88.015802}
}

@article{lopesnpa,
title = {Broken SU(6) symmetry and massive hybrid stars},
journal = {Nucl. Phys. A},
volume = {1009},
pages = {122171},
year = {2021},
issn = {0375-9474},
doi = {https://doi.org/10.1016/j.nuclphysa.2021.122171},
url = {https://www.sciencedirect.com/science/article/pii/S0375947421000361},
author = {Luiz L. Lopes and Débora P. Menezes},
}

@article{lopesPRD,
title = {Baryon coupling scheme in a unified SU(3) and SU(6) symmetry formalism},
  author = {Lopes, Luiz L. and Marquez, Kauan D. and Menezes, D\'ebora P.},
  journal = {Phys. Rev. D},
  volume = {107},
  issue = {3},
  pages = {036011},
  numpages = {16},
  year = {2023},
  month = {Feb},
  publisher = {American Physical Society},
  doi = {10.1103/PhysRevD.107.036011},
  url = {https://link.aps.org/doi/10.1103/PhysRevD.107.036011}
}

@Book{Sakuraibook,
  author    = "J. J. Sakurai",
  title     = "Modern Quantum Mechanics",
  publisher = "Addison Wesley Longman",
  year      = "1994"
}

@article{BETHE1974,
title = {Dense baryon matter calculations with realistic potentials},
journal = {Nucl. Phys. A},
volume = {230},
pages = {1-58},
year = {1974},
issn = {0375-9474},
doi = {https://doi.org/10.1016/0375-9474(74)90528-4},
author = {H.A. Bethe and M.B. Johnson},
}

@article{GLENDENNING1982,
title = {The hyperon composition of neutron stars},
journal = {Phys. Lett. B},
volume = {114},
pages = {392-396},
year = {1982},
issn = {0370-2693},
doi = {https://doi.org/10.1016/0370-2693(82)90078-8},
url = {https://www.sciencedirect.com/science/article/pii/0370269382900788},
author = {N. K. Glendenning},
}

@article{Sedrakian2023,
title = {Heavy baryons in compact stars},
journal = {Prog.  Part.  Nucl. Phys.},
volume = {131},
pages = {104041},
year = {2023},
issn = {0146-6410},
doi = {https://doi.org/10.1016/j.ppnp.2023.104041},
url = {https://www.sciencedirect.com/science/article/pii/S0146641023000224},
author = {A. Sedrakian and J-J. Li and F. Weber},
}

@article{Choro2024,
  title = {Momentum dependence of in-medium potentials: A solution to the hyperon puzzle in neutron stars},
  author = {Chorozidou, Arsenia and Gaitanos, Theodoros},
  journal = {Phys. Rev. C},
  volume = {109},
  issue = {3},
  pages = {L032801},
  numpages = {5},
  year = {2024},
  month = {Mar},
  publisher = {American Physical Society},
  doi = {10.1103/PhysRevC.109.L032801},
  url = {https://link.aps.org/doi/10.1103/PhysRevC.109.L032801}
}
%%%%%%%%%%%%%%%%%%%%%%%%%%%%%

\end{document}